\documentclass[preprint]{elsarticle}
\usepackage{graphicx,url,hyperref,microtype,subcaption,graphbox,amsmath}
\usepackage{graphicx,mathptmx,latexsym,amsmath,amssymb,bm,wrapfig}
\usepackage{latexsym,courier,upgreek,xcolor,rotating,ulem}
\usepackage[onehalfspacing]{setspace}
\hypersetup{colorlinks,citecolor=blue,linkcolor=red,urlcolor=green}
\usepackage[displaymath,mathlines]{lineno}
\usepackage[margin=2cm]{geometry}
\modulolinenumbers[5]
\biboptions{numbers,sort&compress}

\begin{document}

\begin{frontmatter}
  
  \title{The Co-Moving Velocity in Immiscible Two-Phase Flow in Porous Media}
  
  \author[add1]{Subhadeep Roy}
  \ead{subhadeeproy03@gmail.com}
  
  \author[add1]{H\r{a}kon Pedersen}
  \ead{hakon.pedersen@ntnu.no}
  
  \author[add1,add2]{Santanu Sinha}
  \ead{santanu.sinha@ntnu.no}

  \author[add1]{Alex Hansen}
  \ead{alex.hansen@ntnu.no}
    
  \address[add1]{PoreLab, Department of Physics, Norwegian University of Science and Technology, N-7491 Trondheim, Norway.}
  \address[add2]{Beijing Computational Science Research Center, 10 East Xibeiwang Road, Haidian District, Beijing 100193, China.}

\begin{abstract}
We present a continuum (i.e., an effective) description of immiscible two-phase flow in porous media characterized by 
two fields, the pressure and the saturation. Gradients in these two fields are the driving forces that move the 
immiscible fluids around.  The fluids are characterized by two seepage velocity fields, one for each fluid. Following
Hansen et al.\ (Transport in Porous Media, 125, 565 (2018)), we construct a two-way transformation between the velocity
couple consisting of the seepage velocity of each fluid, to a velocity couple consisting of the average seepage velocity 
of both fluids and a new velocity parameter, the co-moving velocity.  The co-moving velocity is related but not equal to 
velocity difference between the two immiscible fluids.  The two-way mapping, the mass conservation equation and the 
constitutive equations for the average seepage velocity and the co-moving velocity form a closed set of equations that
determine the flow.  There is growing experimental, computational and theoretical evidence that constitutive equation 
for the average seepage velocity has the form of a power law in the pressure gradient over a wide range of capillary numbers.
Through the transformation between the two velocity couples, this constitutive equation may be taken directly into 
account in the equations describing the flow of each fluid. This is e.g., not possible using relative permeability theory. 
By reverse engineering relative permeability data from the literature, we construct the constitutive equation for the 
co-moving velocity. We also calculate the co-moving constitutive equation using a dynamic pore network model over a
wide range of parameters, from where the flow is viscosity dominated to where the capillary and viscous forces compete. 
Both the relative permeability data from the literature and the dynamic pore network model give the same very simple functional 
form for the constitutive equation over the whole range of parameters.         
\end{abstract}

\date{\today}
  
  
\end{frontmatter}


\makeatletter
\def\ps@pprintTitle{%
 \let\@oddhead\@empty
 \let\@evenhead\@empty
 \def\@oddfoot{}%
 \let\@evenfoot\@oddfoot}
\makeatother

\section{Introduction}
\label{intro}

When two immiscible fluids compete for the same pore space, we are 
dealing with immiscible two-phase flow in porous media \cite{b88}.    
A holy grail in porous media research is to find a proper description of
immiscible two-phase flow at the continuum level, i.e., at scales where
the porous medium may be treated as a continuum.  Our understanding of 
immiscible two-phase flow at the pore level is increasing at a very high 
rate due to advances in experimental techniques combined with an explosive 
growth in computer power \cite{b17}.  Still, the gap in scales between
the physics at the pore level and a continuum description remains huge and
the bridges that have been built so far across this gap are either complicated to
cross or rather rickety.  To the latter class, we find the still dominating 
theory, first proposed by Wyckoff and Botset in 1936 \cite{wb36} and with an 
essential amendment by Leverett in 1940 \cite{l40}, namely relative permeability 
theory.  The basic idea behind this theory is the following: Put yourself in the
place of one of the two immiscible fluids.  What does this fluid see?  It sees
a space in which it can flow limited by the solid matrix of the porous medium, 
but also by the other fluid.  This reduces its mobility in the porous
medium by a factor known as the relative permeability for that fluid.  
And here is the rickety part: this reduction of available space --- 
expressed through the saturation ---  is assumed to be the {\it only\/} parameter 
affecting the reduction factor or relative permeability. This is a very strong 
statement and clearly does not take into account that the distribution and shape 
of the immiscible fluid clusters will depend on how fast the fluids are flowing, 
and that these two factors affect the reduction of the permeability.  Still, 
in the range of flow rates relevant for many industrial applications, 
this assumption works pretty well. It therefore remains the essential work 
horse for practical applications.

Thermodynamically Constrained Averaging Theory (TCAT) \cite{hg90,hg93,hg93b,nbh11,gm14} 
is a very different approach to immiscible two-phase flow problem.  The TCAT approach is 
generic and not particular to two-fluid flow problems. It is based on 
thermodynamically consistent definitions made at the macro-scale based on volume averages 
of pore-scale thermodynamic quantities. Closure relations are then formulated
at the macro-scale along the lines of the homogenization approach of Whittaker \cite{w86}.  
 A key advantage of TCAT is that all quantities are explicitly defined in terms of 
pore-scale quantities. For example, the pressure that appears in Darcy's law would 
be formally defined as a volume average of the pore-scale pressure field. A key 
disadvantage of TCAT is that very many averaged variables are produced, and many 
complicated assumptions are needed to derive useful results. 

Another development somewhat along the same lines, based on non-equilibrium 
thermodynamics uses Euler homogeneity to define the up-scaled pressure.  
From this, Kjelstrup et al.\ derive constitutive equations for the flow 
\cite{kbhhg19,kbhhg19b}.

Another class of theories is based on detailed and specific assumptions 
concerning the physics involved. An example is Local Porosity Theory
\cite{hb00,h06a,h06b,h06c,hd10,dhh12}.  Another is DeProf (Decomposition in
Prototype Flow) theory which is a fluid mechanical model combined with 
non-equilibrium statistical mechanics based on a classification scheme of 
fluid configurations at the pore level \cite{vcp98,v12,v18}.

Recent work \cite{hsbkgv18,rsh20} has explored a new approach to immiscible 
two-phase flow in porous media based on Euler homogeneity. It provides a 
transformation from the seepage velocity of each fluid to another pair of fluid 
velocities, the average seepage velocity and the {\it co-moving\/} velocity. 
The co-moving velocity, as we shall see, is a velocity parameter that appear 
as a result of the Euler scaling assumption, which is not associated with
any material transport.  The transformation is reversible: knowing the average seepage velocity and
the co-moving velocity, one can determine the seepage velocity of each fluid.
It is the aim of the present work to develop this approach further, especially with
respect to the co-moving velocity. 

A little more than a decade ago, Tallakstad et al.\ \cite{tkrlmtf09,tlkrfm09} injected
simultaneously air and a glycerol-water mixture into a glass-bead filled Hele-Shaw cell
measuring the pressure drop across it as a function of the combined flow rate of the two
fluids, finding a power law relation between them.  Aursj{\o} et al. \cite{aetfhm14} 
repeated the Tallakstad et al.\ experiment, but this time with two incompressible fluids, 
finding the same power law dependency, but with a somewhat different power law exponent.  
The power law relation between pressure difference and flow rate, which corresponds to the 
local average seepage velocity depending on the local pressure gradient to a power when gradients in 
the saturation are negligible, has since been reported by other groups, 
\cite{sh17,glbb20,zbglb21}. This includes finding that the power law regime exists only
in a finite range of pressure gradients; at smaller or larger gradients the relation is
linear. Computational and theoretical approaches to understanding this behavior have 
followed the experimental findings, see 
\cite{sh12,shbk13,xw14,ydkst19,rsh19a,rsh19b,lhrt21,fsrh21}. 

This non-linear constitutive law for the average seepage velocity is a reflection of the
behavior of each of the two immiscible fluids.  The Euler approach of Hansen et al.\
\cite{hsbkgv18,rsh20} makes it possible to transform this constitutive law describing the
local average seepage velocity as a function of the local driving forces into constitutive laws for 
each of the two fluids.  However, this hinges on providing a constitutive law for the 
co-moving velocity.  

We will in this paper develop a constitutive equation for the co-moving velocity
under the assumption that gradients in the saturation may be neglected.  
Together with the constitutive equation for the average 
velocity, we then have a complete description of the flow as long as there are
no saturation gradients.

Generalizing our results to when there are saturation gradients will be the subject of
future investigations. 

We investigate the constitutive equation for the co-moving velocity using two approaches.  
The first one is to use experimental relative permeability data from the literature 
to construct the constitutive equation for the co-moving velocity.  Since the relative 
permeability approach obeys the Euler homogeneity assumption, it is possible to express
the co-moving velocity in terms of the relative permeabilities.  This opens up for 
{\it reverse engineering\/} the experimental data which have been cast in terms of 
relative permeability curves in order to construct the co-moving velocity.  

It should be noted here that this reverse engineering of the data does {\it not\/} rely on
the relative permeability constitutive equations being accurate or even correct.  It simply 
consists of translating the data that have been cast in the form of relative permeability data into seepage
velocity data that in turn allow us to construct the co-moving velocity.              

The second approach is based on a dynamic pore network model  \cite{jh12} first introduced by 
Aker et al.\ \cite{amhb98} and then later refined \cite{gvkh19,gwh19}. A review of 
the model was recently published by Sinha et al.\ \cite{sgvh20}.  It allows us to emulate
closely the experiments of Tallakstad et al.\ \cite{tkrlmtf09,tlkrfm09}, e.g. reproducing
the power law dependence of the flow rate on the pressure drop \cite{sh12}.    

The constitutive law for the co-moving velocity turns out to be 
surprisingly simple, see equation (\ref{relperm-200}).  
The reason for this remains an open question.

The main body of the paper is divided into three sections.  The first one, 
Section \ref{secThermo}, reviews the Euler homogeneity approach to 
immiscible two-phase flow in porous media \cite{hsbkgv18,rsh20}. The Section
starts by laying the groundwork for the theory by defining central variables.

In Subsection \ref{hysteresis}, we address central questions concerning these
variables: Are they at all possible to define or will they be swamped by 
fluctuations?  Is it possible to see them as {\it state variables,\/} that is,
variables that describe the flow there and then without depending on the history
of the system? Can we still deal with these variables when there is hysteresis? 
After this discussion, we go on to describe in Subsection \ref{homogeneity} 
the consequences of the volumetric flow rate being an Euler homogeneous function 
in the area over which the volume is measured.  
We then go on in Subsection \ref{closed} describe how the equations of the previous
subsection together with constitutive equations for the local average seepage velocity
and the local co-moving velocity form a closed set of equations that determine 
the local seepage velocities of the fluids, the local saturation and the local
pressure field.  
In Subsection \ref{interpreting} we give a physical interpretation of the meaning
of the co-moving velocity. 
The next Section \ref{expResults} we turn to analyzing experimental data from
the literature that allow us to reconstruct the co-moving velocity.  
We start this section by describing (Subsection \ref{RelPermEuler}) how relative 
permeability theory may be cast in the language of the Euler scaling approach
of Section \ref{secThermo}. In this way, we relate the relative permeabilities
to the co-moving velocity.  We emphasize yet again that 
this does not imply that relative permeability theory is correct.  Rather, the 
assumption is: If we assume the central equations of relative permeability
theory, then the co-moving velocity could be expressed in terms of relative
permeability curves, see equations (\ref{RelPereq05}) or (\ref{relperm-1}).    
The next Subsection \ref{AnalysisRelPerm} present our analysis of 
different relative permeability data sets including the reconstructed
co-moving velocity, see equation (\ref{relperm-200}).  This is the main 
result in this paper. 
Section \ref{poreNetwork} focuses on using a dynamic pore network model
to calculate the co-moving velocity.  Subsection \ref{vp_sw} details how
we extract the average seepage velocity and then the co-moving velocity from the numerical
data generated by the model. We then fit the data to the form (\ref{relperm-200})
in Subsection \ref{older_vm}, finding excellent agreement. 
Hansen et al.\ \cite{hsbkgv18} presented the co-moving velocity gotten by 
the dynamic pore network model, but using a different set of variables
than we use here.  Subsection \ref{a_b_deltaP} discusses the relation between
the functional form we find for the co-moving velocity here and the one 
found in Hansen et al.
We have earlier in this introduction described the work of 
Tallakstad et al.\ \cite{tkrlmtf09,tlkrfm09} and subsequent workers,
where a non-linear relation between average seepage velocity and pressure gradient
was uncovered. In Subsection \ref{vp_nonlinear} we report on what happens
to the co-moving velocity when the flow is in the regime.  The interesting
answer is it does not change character.
We then go on to investigate in Subsection \ref{limits} 
what happens to the co-moving velocity
when the wetting saturation or the non-wetting saturation falls below 
the threshold for two-phase flow. We see a change in the coefficients 
describing the co-moving velocity, but not its functional form when
the wetting saturation falls below the two-phase flow threshold.  However,
no such jump is seen at the other threshold.  We note that there is
hysteresis associated with the low wetting saturation threshold but not 
with the low non-wetting saturation threshold \cite{kh06}. 
Lastly in this section, we discuss the effect of changing the viscosity ratio of the two immiscible
fluids on the co-moving velocity, see Subsection \ref{viscosity}.   
Finally, we draw our conclusions in Section \ref{discussion}.

\section{Euler scaling approach}
\label{secThermo}

We consider in the following two incompressible and immiscible fluids, 
one of which more wetting with respect to the pore matrix than the other.  We will refer
to the first fluid as the wetting fluid and the second one the non-wetting fluid. 
The viscosity of the wetting fluid is $\mu_w$ and of the non-wetting fluid $\mu_n$.

We consider a porous medium at a scale where it may be viewed as a continuum.  
This is a scale that is much larger than the pore scale. Whereas at the pore
scale concepts such as fluid clusters, interfaces and wetting are central, 
they are not useful at the continuum scale.  Rather, different concepts, and hence 
variables, should be --- and to some degree are --- used.  This is the viewpoint 
will retain throughout this section.    

This viewpoint has consequences. In this continuum limit, the pores are essentially 
infinitely small, and so are the fluid interfaces in the pores.  Hence, it is no longer 
fruitful to view the problem as the flow of two immiscible fluids since the key notions 
that belong to such a description all are closely related to pore-scale concepts.  
Rather, the two immiscible fluids may be seen acting as a {\it single\/} fluid whose 
rheological properties --- for example the effective viscosity --- is controlled by 
two variables, the pressure $P$ and the wetting saturation $S_w$.   

There are two driving forces in the continuum limit description that get this single 
fluid to move: spatial gradients in the pressure, $\nabla P$ and the saturation, $\nabla S_w$. 
The latter driving force has its origin at the pore level in capillary forces.  We may express 
this driving force in terms of a field with the dimensions of pressure, $P_c$, which depends 
on the saturation, so that $\nabla P_c(S_w)=(dP/dS_w)\nabla S_w$.  In relative 
permeability theory, we would call $P_c$ the dynamic capillary pressure field.  

It is necessary to describe the single fluid using {\it two\/} velocity fields. 
This is a reflection of the saturation not being transported at the same 
velocity as the fluid itself.  We name the velocity field that transports the
fluid $\vec v_p$ and the velocity field that transports the wetting saturation 
$\vec v_w$,
\begin{equation}
\label{euler-1}
\phi\ \frac{\partial S_w}{\partial t}=\nabla\cdot \vec v_w \phi S_w\;,
\end{equation}
where $\phi$ is the porosity field and $t$ is time.

We define the porosity field as follows: We may associate with each point in the 
porous medium a Representative Elementary Volume (REV) which is very large compared
to the pore scale, but small compared to continuum scale. The porosity of a given
point is then the pore volume of REV divided by the volume of the REV. We note
that there might be structure in the porous medium at the continuum scale, so
that the porosity field may vary spatially, generating a non-zero gradient 
$\nabla\phi$. 

We also define a Representative Elementary Area (REA) \cite{bb12}. We pick a point 
in the porous medium. There will be a stream line associated with the velocity field 
$\vec v_p$ at that point.  We place a plane of area $A$ orthogonal to the stream line centered 
at the point.  We assume that the plane is small enough so that the other 
stream lines passing through the plane all are essentially parallel to the first one.  
We also assume that the plane is small enough for the porous medium to be homogeneous 
over the size of the plane with respect to porosity and permeability. This is the REA. 

This allows us to define a transverse pore area    
\begin{equation}
\label{eqn0.01}
A_p=\phi A\;.
\end{equation} 
The transverse pore area is the area of the REA that cuts through the pores. 

The transverse pore area $A_p$ may be split into a transverse wetting fluid 
area $A_w$ and a transverse non-wetting fluid area $A_n$.  We mean by $A_w$ 
the area of the plane covered by the wetting fluid and $A_n$ the area covered 
by the non-wetting fluid. We have that   
\begin{equation}
\label{eqn5}
A_p=A_w+A_n\;.  
\end{equation}  
The wetting and non-wetting saturations $S_w$ and $S_n$ may be expressed as
\begin{equation}
\label{eqn3}
S_w=\frac{A_w}{A_p}\;,  
\end{equation}
and
\begin{equation}
\label{eqn4}
S_n=\frac{A_n}{A_p}\;,  
\end{equation}
so that 
\begin{equation}
\label{eqn4-1}
S_w+S_n=1\;.
\end{equation}

There is a volumetric flow rate $Q_p$ passing through the plane which may be 
decomposed into a volumetric flow rate for the wetting fluid, $Q_w$, and a 
volumetric flow rate for the non-wetting fluid $Q_n$.  We have that
\begin{equation}
\label{eqn1}
Q_p=Q_{w}+Q_{n}\;.  
\end{equation}
This allows us to define three velocities,
\begin{equation}  
\label{eqn5.1}
v_w=\frac{Q_w}{A_w}\;,
\end{equation}
and 
\begin{equation}  
\label{eqn5.2}
v_n=\frac{Q_n}{A_n}\;,
\end{equation}
and
\begin{equation}  
\label{eqn5.4}
v_p=\frac{Q_p}{A_p}=
\frac{A_w}{A_p}\ \frac{Q_w}{A_w}+\frac{A_n}{A_p}\ \frac{Q_n}{A_n}=S_wv_w+S_nv_n\;.
\end{equation}
These are the {\it seepage velocities.\/}  We will refer to $v_p$ as the 
{\it average seepage velocity\/} in the following.

We may note here that since we are assuming the fluids to be incompressible, 
it makes no difference whether we define the seepage velocities of each fluid
with respect to volume flow or mass flow.  However, the average seepage velocity $v_p$,
defined in equation (\ref{eqn5.4}) will be different if averaged with respect to
mass rather than volume.  The formalism we are about to develop in Section 
\ref{homogeneity} and onwards, could have been done using this averaging instead. 
We have, however, decided to stick with volume averaging. 

\subsection{Fluctuations, state variables and hysteresis}
\label{hysteresis}

We will in the following sections treat the variables we have just defined as functions of each other, 
even to the point of taking derivatives. In this section, we pose the question of whether this 
is at all possible. There are three aspects we need to address in this context:
The first one concerns {\it fluctuations.\/}  If the variables we consider fluctuate strongly, it is not 
possible to find functional relations between them. The second aspect is the question of whether the 
variables we measure depend only on the flow there and then or whether they in addition depend on 
the history of the flow.  If the former is true, we are dealing with {\it state variables.\/}
The third aspect concerns the possibility of these variables being multi-valued.  That is, there 
is {\it hysteresis.\/}  Is the analysis we present still valid when there is hysteresis? 

{\it Fluctuations:\/} Self-averaging is an important property of fluctuating systems.  A self-averaging
system is one where the relative strength of the fluctuations shrinks with increasing size of the 
system.  If this is so, the variables attain well-defined values and functional relations between
them may be sought.  

To give an example, this is precisely the situation when thermodynamics is used to describe a gas.  The 
more molecules it consists of, the more well defined the macroscopic thermodynamic variables and their 
relations are.  We note, however, that in such systems there is one exception: At critical points, the 
fluctuations dominate and self-averaging is lost \cite{ah96}.  

An important feature of flooding processes, slow or fast, is that they typically 
generate fractal injection patterns \cite{ffh22}. These patterns, like the fluctuations near 
critical points, are typically not self-averaging.  However, there will always be a largest 
length scale above which the process does not produce fractals.  Here, self-averaging sets in.  In the 
continuum limit --- which is what we consider here --- we are surely above this scale.

It should be noted that there is not a one-to-one correspondence between the fluid configurations
and the values of the macroscopic variables. Rather, typically there are many fluid configurations giving rise
to the same values for the macroscopic variables.  This is not a problem as it is the macroscopic variables
that are measured, not the underlying fluid configurations. 

One may then ask oneself, does this mean that the theory being developed here is untestable on 
small systems such as those that can be modeled using computational method such 
as the lattice Boltzmann method or dynamic pore network models since we can never reach sufficient system
sizes for the fluctuations to be small enough?  The answer is no as one may use time averaging to emulate 
size.  In fact, Kjelstrup et al.\ \cite{kbhhg19} report that around 100 links are enough to define a REV in
the dynamic pore network model \cite{sgvh20} we explore further on in this paper. 

{\it State variables:\/} Steady-state flow of immiscible fluids in a porous medium needs to be carefully defined.
We have settled on the following: It is a flow where the macroscopic variables have values (measured in 
practice as gliding averages over time) that do not drift in any direction.  This does not preclude fluid 
clusters moving, merging and breaking up.  In three-dimensional flow, one may have that both fluid phases
percolate.  If the flow then is not too fast, the fluid interfaces will not move.  However, when there is
no percolation of either phase, which is typically easier to obtain in two-dimensional systems, the clusters
will exhibit a rich dynamics.  

Erpelding et al.\ \cite{esthfm13} studied experimentally and computationally such a two-dimensional system.
Their experimental set-up consisted of a two-dimensional (42 cm $\times$ 85 cm) Hele-Shaw cell filled with 
immobilized 1 mm glass beads.  Along one of the short edge, two immiscible fluids (a water-glycerol mixture 
and air) were injected simultaneously through 15 alternating injection points at constant rates. The opposite
edge of the Hele-Shaw cell was left open, and the two orthogonal sides were both sealed.  Hence, there would
be a flow across the cell from the injection points in the direction of the open edge.  Some distance from the
injection points in the flow direction, the fluids would mix sufficiently to create a mixture of
fluid clusters that when averaged over time would be homogeneous.  

This system would be set up at a given flow rate and a number of variables were measured.  The flow 
rate would then be raised and new values for the variables would be measured. Then, the flow rate would
revert to the original value and the variables measured anew.  The variables would attain the values they
had before the flow rate was raised. The flow is {\it history independent\/} in the language of Erpelding et 
al.\ \cite{esthfm13}, and the macroscopic variables describing it would then be {\it state variables.\/} 
They would characterize the flow there and then, and not depend on the history of the flow.  

{\it Hysteresis:\/} There is the hysteresis caused by the difference between first and secondary 
flooding \cite{b17}.  Typically at low injection rates, the system will remember its history and 
the values for the macroscopic variables will be different when the first and second time one 
floods the system.   

There is, however, also another kind of hysteresis which is related to the study of Erpelding et al.\
\cite{esthfm13}.  Modeling the Hele-Shaw system, Knudsen and Hansen \cite{kh06} studied the wetting 
fractional flow as function of wetting saturation under steady-state conditions using a dynamic pore 
network model.  They found that there are two transitions between two-phase flow and single-phase flow 
when the saturation is the control parameter.  The transition between only the non-wetting fluid moving 
at low saturation to both fluids moving at higher saturation does not show any hysteresis with respect 
to which way one passes through the transition.  However, the other transition between only the 
wetting fluid moving at high saturation and both fluids moving at lower saturation does show a 
strong hysteresis. This is depicted in Figure 2 in Reference \cite{kh06}.  This hysteresis, we believe, 
is caused by this transition being related to a first order (or spinodal) phase transition.  

Hysteretic behavior is a signal that the macroscopic state variables are multi-valued, signaling
--- of course --- that the underlying microscopic physics has more than one stable mode.  
Hysteresis is far from uncommon in physics.  In fact, it is a defining property of 
first order phase transitions.  There are no principal problems manipulating multi-valued 
functions, for example taking their derivatives as long as one does not mix up the branches.  

\subsection{Homogeneity of $Q_p$ and consequences thereof}
\label{homogeneity}

In the following we review the central arguments in  \cite{hsbkgv18}.

The volumetric flow rate $Q_p$ across the REA is a homogeneous function of order one 
in the transverse area variables $A_w$ and $A_n$.  That is,  
\begin{equation}
\label{eqn0.100}
Q_p(\lambda A_w,\lambda A_n)=\lambda Q_p(A_w,A_n)\;,
\end{equation}
where $\lambda$ is a scale factor. Taking the derivative 
with respect to $\lambda$ and setting $\lambda=1$ in this expression, 
we get, 
\begin{equation}
\label{eqn0.101}
Q_p(A_w,A_n)=A_w\left(\frac{\partial Q_p}{\partial A_w}\right)_{A_n}+
A_n\left(\frac{\partial Q_p}{\partial A_n}\right)_{A_w}\;.
\end{equation}
Dividing $Q_p$ in this equation by the transverse pore area $A_p$, we get
\begin{align}
\label{eqn0.3}
v_p = S_w\left(\frac{\partial Q_p}{\partial A_w}\right)_{A_n} + 
S_n\left(\frac{\partial Q_p}{\partial A_n}\right)_{A_w} =S_w \hat{v}_w + S_n\hat{v}_n\;.  
\end{align}
where 
\begin{equation}
\label{eqn0.1}
\hat{v}_w=\left(\frac{\partial Q_p}{\partial A_w}\right)_{A_n}\;,
\end{equation}
and
\begin{equation}
\label{eqn0.2}
\hat{v}_n=\left(\frac{\partial Q_p}{\partial A_n}\right)_{A_w}\;,
\end{equation}
are the {\it thermodynamic velocities.\/} They differ from
the seepage velocities (\ref{eqn5.1}) and (\ref{eqn5.2}) as we shall see, 
this in spite of $v_p$ being given by both (\ref{eqn5.4}) and (\ref{eqn0.3}).

We may express the two thermodynamic velocities $\hat{v}_w$ and $\hat{v}_n$
in terms of the average seepage velocity $v_p$. In order to do so,
we change our control variables from $(A_w,A_n)$ to $(A_p,S_w)$. 
We use equations (\ref{eqn3}) and (\ref{eqn4}) and the chain rule to derive
\begin{eqnarray}
\label{eqn0.1000}
\left(\frac{\partial}{\partial A_w}\right)_{A_n}
&=&\frac{S_n}{A_p}\ \left(\frac{\partial }{\partial S_w}\right)_{A_p}
+\left(\frac{\partial }{\partial A_p}\right)_{S_w}\;,
\end{eqnarray}
and 
\begin{equation}
\label{eqn0.1001}
\left(\frac{\partial}{\partial A_n}\right)_{A_w}
=-\frac{S_w}{A_p}\ \left(\frac{\partial }{\partial S_w}\right)_{A_p}
+\left(\frac{\partial }{\partial A_p}\right)_{S_w}\;.
\end{equation}
We now combine these two equations with the definitions (\ref{eqn0.1}) 
and (\ref{eqn0.2}), and use $Q_p=A_pv_p$, i.e.\ equation (\ref{eqn5.4}), 
to find
\begin{equation}
\label{eqn0.3000}
\hat{v}_w=v_p+S_n\frac{dv_p}{dS_w}\;,
\end{equation}
and
\begin{equation}
\label{eqn0.3001}
\hat{v}_n=v_p-S_w\frac{dv_p}{dS_w}\;.
\end{equation} 
This is a remarkable result in that $\hat{v}_w$ and $\hat{v}_n$ are fully 
determined by $v_p$ and its derivative with respect to $S_w$.  In other words,
it is enough to know $v_p(S_w)$ to determine {\it both\/} $\hat{v}_w$ and $\hat{v}_n$. 

From equations (\ref{eqn5.4}) and (\ref{eqn0.3}), we have that
\begin{equation}
\label{eqn5002}
S_w v_w+S_n v_n =S_w \hat{v}_w+S_n\hat{v}_n\;.
\end{equation}
The most general relation between between $(\hat{v}_w,\hat{v}_n)$ and 
$(v_w,v_n)$ is given by the pair of equations 
\begin{equation}
\label{eqn0.4}
\hat{v}_{w}=v_{w}+S_{n}v_m\;,  
\end{equation}
and 
\begin{equation}
\label{eqn0.5}
\hat{v}_{n}=v_{n}-S_{w}v_m\;,  
\end{equation}  
where a new velocity function $v_m$ has been introduced.  This is the 
{\it co-moving velocity.\/} 

Equations (\ref{eqn0.4}) and (\ref{eqn0.5}) {\it define\/} the co-moving velocity.  The
co-moving velocity provides the link between the seepage velocities and the 
thermodynamic velocities. 
  
We combine the two equations (\ref{eqn0.4}) and 
(\ref{eqn0.5}) with equations (\ref{eqn0.3000}) and (\ref{eqn0.3001}), 
to find
\begin{equation}
\label{eqn0.3010}
v_w=v_p+S_n\left(\frac{dv_p}{dS_w}-v_m\right)\;,
\end{equation}
and
\begin{equation}
\label{eqn0.3011}
v_n=v_p-S_w\left(\frac{dv_p}{dS_w}-v_m\right)\;.
\end{equation}
Thus, we have expressed the seepage velocity for each fluid $v_w$ and $v_n$ in
terms of the average seepage velocity $v_p$ and the co-moving velocity $v_m$.  
This is in contrast to the thermodynamic velocities $\hat{v}_w$ and $\hat{v}_n$
where only the average seepage velocity $v_p$ was needed, see equations 
(\ref{eqn0.3000}) and (\ref{eqn0.3001}).

We may see equations (\ref{eqn0.3000}) and (\ref{eqn0.3001}) as a mapping
$(v_p, v_m)\to(v_w,v_n)$. The couple $(v_p,v_m)$ contains the same information
as the couple $(v_w,v_n)$.    

The co-moving velocity was defined in equations (\ref{eqn0.4}) and (\ref{eqn0.5}). We
may express it explicitly by solving (\ref{eqn0.3010}) and (\ref{eqn0.3011}) with 
respect to $v_m$, finding
\begin{equation}
\label{eqn19}
v_m=\frac{dv_p}{dS_{w}}+\left(v_n-v_w\right)\;. 
\end{equation}
If we now take the derivative of equation (\ref{eqn5.4}) with respect to $S_w$
and combine the resulting equation with equation (\ref{eqn19}), we find 
\begin{equation}
\label{eqn20}
v_m = S_w\frac{dv_w}{dS_w} + S_n\frac{dv_n}{dS_w}\;.  
\end{equation}
We may take either of equations (\ref{eqn19}) and (\ref{eqn20}) as alternative 
definitions of the co-moving velocity.  

For clarity, we now display equations (\ref{eqn5.4}) and (\ref{eqn20}) together
as follows:
\begin{eqnarray}
v_p=S_wv_w+S_nv_n\;,\nonumber\\
v_m=S_wv'w+S_nv'_n\;,\nonumber
\end{eqnarray} 
where we have used the notation $v'_w=dv_w/dS_w$ and $v'_n=dv_n/dS_w$.  These
two equations, (\ref{eqn5.4}) and (\ref{eqn20}), give us the reverse transformation
$(v_w,v_n)\to(v_p,v_m)$.   

\subsection{Closed set of equations}
\label{closed}

We defined the Representative Elementary Area in Section \ref{secThermo}. Its size 
was determined by the largest transverse area over which the streamlines could be
regarded as parallel. On larger scales, the stream lines form patterns that reflect
the structure and boundaries of the porous medium; e.g., a reservoir. In this Section,
we construct a closed set of equations that determine the flow at these scales based on the
formalism constructed in the previous Section, conservation laws and constitute equations.            

The plane with area $A$ we introduced in the preceding sub-section was oriented
orthogonally to the stream line for $v_p$ at the point it sits.  We may orient it
differently generating the same equations, but with the velocities now being 
components relative to the axis of the new plane.  This makes it possible to
express the equations in terms of vectors. 

The fluids are incompressive so that
\begin{equation}
\label{eqn0.4500}
\nabla\cdot \phi\vec v_p=0\;.
\end{equation}
We have here assumed that the porosity may not be spatially uniform.
The continuity equation for the wetting saturation, $S_w$, equation (\ref{euler-1}),
may be combined with the vector version of equation (\ref{eqn0.3010}) to give
\begin{equation}
\label{euler-10}
\phi\ \frac{\partial S_w}{\partial t}=\nabla\cdot 
\left[\vec v_p+S_n\left(\frac{d\vec v_p}{dS_w}-\vec v_m\right)\right] \phi S_w\;.
\end{equation}

These two continuity equations must be supplied with two constitutive equations
\begin{equation}
\label{euler-11}
\vec v_p=\vec v_p(S_w,\nabla S_w,\nabla P)\;,
\end{equation}
and
\begin{equation}
\label{euler-12}
\vec v_m=\vec v_m(S_w,\nabla S_w,\nabla P)\;,
\end{equation}
to produce a closed set of equations that together with the proper boundary and
initial values solves the immiscible two-phase flow problem in the continuum limit.

We note that the non-linear constitutive equation that can be constructed for $\vec v_p$ 
from the observations in \cite{tkrlmtf09,tlkrfm09,aetfhm14,sh17,glbb20,zbglb21}, 
is easily combined with this approach.    

\subsection{Interpreting the co-moving velocity $v_m$}
\label{interpreting}

Let us now pose the question: is $\vec v_m$ transporting anything?  Equations 
(\ref{eqn5.1}), (\ref{eqn5.2}) and (\ref{eqn5.4}) show that there is volumetric transport
associated with the velocities $\vec v_w$, $\vec v_n$ and $\vec v_p$. We will in the 
following show that there is no such transport associated with $\vec v_m$.

We base the discussion that now follows on \cite{rsh20}. We will consider components
rather than vectors.  We introduce the {\it differential transverse area 
distributions\/} $a_p$, $a_w$ and $a_n$. Their meaning is as follows: 
$a_p(v) dv$ is the area covered by fluid, wetting or non-wetting, that 
has a velocity in the interval $[v,v+dv]$. Likewise, $a_w(v) dv$ is the 
area covered by wetting fluid that has a velocity in the 
interval $[v,v+dv]$ and $a_n(v) dv$ is the area covered by non-wetting fluid 
that has a velocity in the interval $[v,v+dv]$.  Hence, we have that
\begin{equation}
\label{euler-13}
A_p=\int_{-\infty}^\infty dv\ a_p\;,
\end{equation}
\begin{equation}
\label{euler-14}
A_w=\int_{-\infty}^\infty dv\ a_w\;,
\end{equation}
and 
\begin{equation}
\label{euler-15}
A_n=\int_{-\infty}^\infty dv\ a_n\;.
\end{equation}

The velocities defined in equations (\ref{eqn5.1}), (\ref{eqn5.2}) and
(\ref{eqn5.4}) are then given by 
\begin{equation}
\label{euler-16}
v_p=\frac{1}{A_p}\int_{-\infty}^\infty vdv\ a_p\;,
\end{equation}
\begin{equation}
\label{euler-17}
v_w=\frac{1}{A_w}\int_{-\infty}^\infty vdv\ a_w\;,
\end{equation}
and 
\begin{equation}
\label{euler-18}
v_n=\frac{1}{A_n}\int_{-\infty}^\infty vdv\ a_n\;.
\end{equation}

The differential transverse areas are essentially velocity histograms,
thus making a connection between the continuum scale and the flow at small scales. 

We may now combine these three equations, (\ref{euler-16}),
(\ref{euler-17}) and (\ref{euler-18}), with equation (\ref{eqn19})
to give
\begin{equation}
\label{euler-19}
v_m=\frac{dv_p}{dS_w}-v_w+v_n=\frac{1}{A_p}
\int_{-\infty}^\infty vdv\ 
\left[\frac{\partial a_p}{\partial S_w}-\frac{a_w}{S_w}+\frac{a_n}{S_n}\right]
=\frac{1}{A_p} \int_{-\infty}^\infty vdv\ a_m\;, 
\end{equation}  
from which we infer
\begin{equation}
\label{euler-20}
a_m(v)=\frac{\partial a_p(v)}{\partial S_w}-\frac{a_w(v)}{S_w}+\frac{a_n(v)}{S_n}\;.
\end{equation} 
This is the co-moving differential transverse area. We now integrate this over all
velocities to find the total co-moving transverse area $A_m$,
\begin{eqnarray}
\label{euler-21}
A_m&=&\int_{-\infty}^\infty dv\ a_m=\frac{d}{dS_w}\int_{-\infty}^\infty dv\ a_p
-\frac{1}{S_w}\int_{-\infty}^\infty dv\ a_w+\frac{1}{S_n}\int_{-\infty}^\infty dv\ a_n\nonumber\\
&=&\frac{dA_p}{dS_w}-\frac{A_w}{S_w}+\frac{A_n}{S_n}=0\;.
\end{eqnarray}
There is no area associated with the co-moving velocity.  As a consequence, there is no
volumetric flux associated with it as
\begin{equation}
\label{euler-22}
Q_m=A_m v_m=0\;.
\end{equation}
Both of these results make sense, since $A_w+A_n=A_p$ (equation (\ref{eqn5}))
and $Q_w+Q_n=Q_p$ (equation (\ref{eqn1})): There is no room for $v_m$ being
associated with any transverse area or with volumetric transport.  We may see the transformation 
$(v_w,v_n)\to(v_p,v_m)$ as a way of partitioning the flow. $(A_w,A_n)$ 
and $(Q_w,Q_n)$ constitute one partitioning, $(A_p,A_m)=(A_p,0)$ and 
$(Q_p,Q_m)=(Q_p,0)$ another.     

Equation (\ref{eqn19}) shows that $v_m$ is related to the relative velocity of 
the two fluids, $v_n-v_w$.  However, the difference velocity, $v_n-v_w$ cannot be
given an interpretation as being part of a partitioning of the flow.

Before we now switch to the structure of the co-moving velocity $v_m$, it is now appropriate
to remind the reader of why the mapping $(v_w,v_n)\rightleftarrows (v_p,v_m)$, that is equations
(\ref{eqn5.4}) and (\ref{eqn20}) for the transformation $(v_w,v_n)\rightarrow (v_p,v_m)$, and
equations (\ref{eqn0.3010}) and (\ref{eqn0.3011}) for the transformation 
$(v_w,v_n)\leftarrow (v_p,v_m)$, is important.  With the non-linear constitute law for $v_p$
being uncovered experimentally, computationally and theoretically 
\cite{tkrlmtf09,tlkrfm09,aetfhm14,sh17,glbb20,zbglb21,sh12,shbk13,xw14,ydkst19,rsh19a,rsh19b,lhrt21,fsrh21},
a theory that can relate this constitutive law to the flow properties of each of the immiscible fluids
is necessary. It is precisely such a theory that we are presenting here.       

\section{Reverse engineering relative permeability data}
\label{expResults}

Our aim is now to {\it reverse engineer\/} experimental data from the literature that have
been presented as relative permeability curves to reconstruct a constitutive equation for the 
co-moving velocity.    

In order to do so, we begin this section by placing relative permeability theory within  
the framework of the Euler homogeneity approach.  This allows us to express the co-moving
velocity $v_m$ in terms of the relative permeabilities.  

It is important to note here that this approach does not hinge on whether the relative 
permeability approach is correct or not.  Rather, we are simply translating the data back
to their origin and from there we construct $v_m$.   

Which relative permeability data sets to choose?  Since we have no preconceived ideas of 
the form of $v_m$ or what controls it, we have more or less randomly picked relative
permeability data sets.  Any other way of picking them would bias the results.  

We note that the relative permeability data are hysteretic.  There is, however, no
problem in taking the derivatives of these curves in order to extract the co-moving 
velocities.  It might be that the co-moving velocities also are hysteretic. At this point,
we do not know.     

\subsection{Relative permeability theory in light of Euler homogeneity}
\label{RelPermEuler} 

Relative permeability theory \cite{wb36} is based on the two constitutive equations, 
\begin{align}\label{RelPereq01}  
\vec v_w = - \displaystyle\frac{K k_{rw}}{\phi S_w\mu_w} \nabla P\;,
\end{align}
and 
\begin{align}\label{RelPereq02}  
\vec v_n = - \displaystyle\frac{K k_{rn}}{\phi S_n\mu_n} \nabla P\;,
\end{align}
when we assume that there are no saturation gradients so that $\nabla P_c=0$ \cite{v18b}. 
Here $K$ is the absolute permeability. The factors $ k_{rw}=k_{rw}(S_w)$ and $k_{rn}= k_{rn}(S_w)$ 
are the wetting and non-wetting relative permeabilities. 

We introduce the plate of area $A$ as in Section \ref{homogeneity} and form the 
volumetric flow rate through it, $Q_p$.  From this we get $v_p$ by using equation 
(\ref{eqn5.4}). Combining this equation with the the relative permeability constitutive
equations, also named the generalized Darcy equations 
(\ref{RelPereq01}) and (\ref{RelPereq02}) gives
\begin{equation}
\label{RelPereq04a}
v_p=- \mu_w v_0\left[\frac{k_{rw}}{\mu_w}+\frac{k_{rn}}{\mu_n}\right]\;,
\end{equation}
where we have introduced a velocity scale which is independent of $S_w$, 
\begin{align}\label{RelPereq06}  
v_0=-\frac{K}{\mu_w\phi}|\nabla P|\;.
\end{align}
We see that this is an Euler homogeneous function of order zero in $A_w$ and $A_n$ 
implying that $Q_p=(A_w+A_n)v_p$ trivially fulfills equation (\ref{eqn0.100}).   
Hence, relative permeability theory obeys all the relations we derive in 
Section \ref{homogeneity}.

We now combine the generalized Darcy equations (\ref{RelPereq01}) and (\ref{RelPereq02})
with equation (\ref{eqn20}) for the co-moving velocity $v_m$.  We find    
\begin{align}\label{RelPereq05}  
v_m = \mu_w v_0\left[\frac{S_w}{\mu_w}\frac{d}{dS_w}
\left(\frac{k_{rw}}{S_w}\right)+\frac{S_n}{\mu_n}
\frac{d}{dS_w}\left(\frac{k_{rn}}{S_n}\right)\right]\;.
\end{align}
We may also write $v_m$ as 
\begin{equation}
\label{relperm-1}
v_m=\frac{dv_p}{dS_w}+\mu_w v_0\left[\frac{k_{rn}}{\mu_n}-\frac{k_{rw}}{\mu_w}\right]\;,
\end{equation}
using equation (\ref{eqn19}). 

\begin{figure}
\centering
\includegraphics[width=0.8\textwidth,clip]{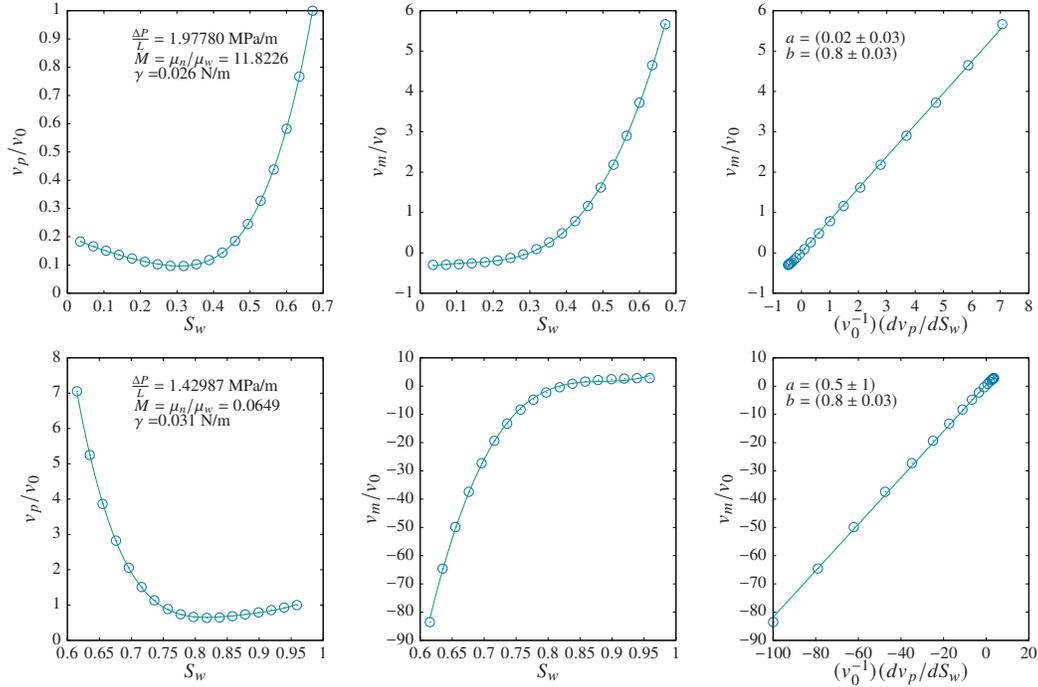}
\caption{Experimental data from Table 3 in Bennion et al.\ \cite{bb05}. Upper row:
Basal Cambrian sandstone, drainage with brine as non-wetting fluid and CO$_2$
as wetting fluid, $\phi = 0.177$, $K = 5.43\times10^{-16}$ m$^2$. 
Lower row: Wabamum carbonate, drainage with CO$_2$ as non-wetting fluid and brine 
as wetting fluid, $\phi = 0.177$, $K = 2.07 \times 10^{-16}$ m$^2$.}
\label{fig:Bennion_experimental}
\end{figure}

\begin{figure}
\centering
\includegraphics[width=0.8\textwidth,clip]{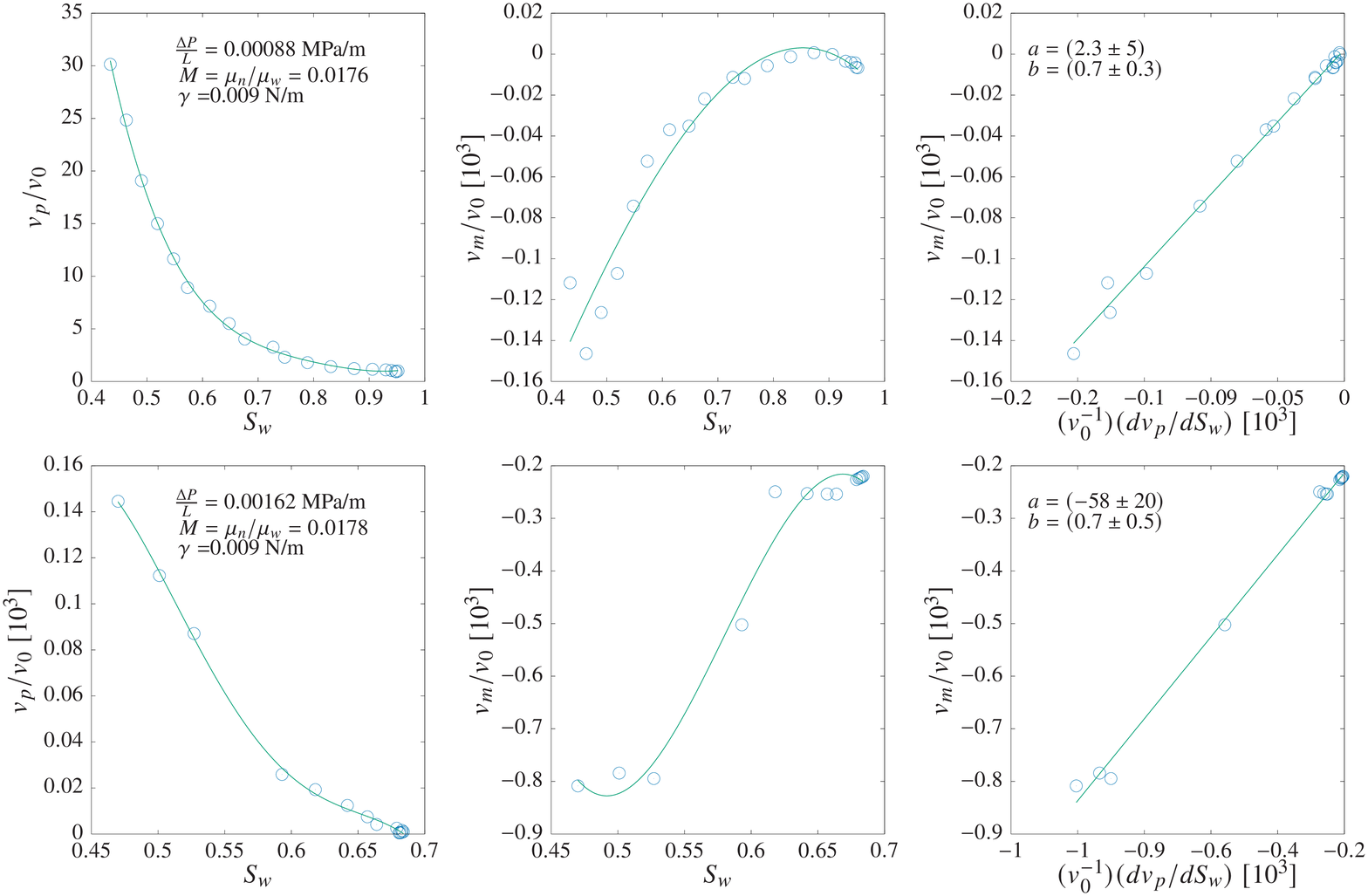} 
\caption{Experimental data from figure 4 in Oak et al.\ \cite{obt90}. 
The process is primary drainage (upper row) and imbibition (lower row) of a 
single experimental run. Both rows: Berea sandstone, natural gas as 
non-wetting fluid and water as wetting fluid, $\phi = 0.193$, 
$K = 2072.49 \times 10^{-16}$ m$^2$. $\phi$ and $\gamma$ were not 
supplied by the authors.}
\label{fig:Oak_experimental_WG}
\end{figure} 

\begin{figure}
\centering
\includegraphics[width=0.8\textwidth,clip]{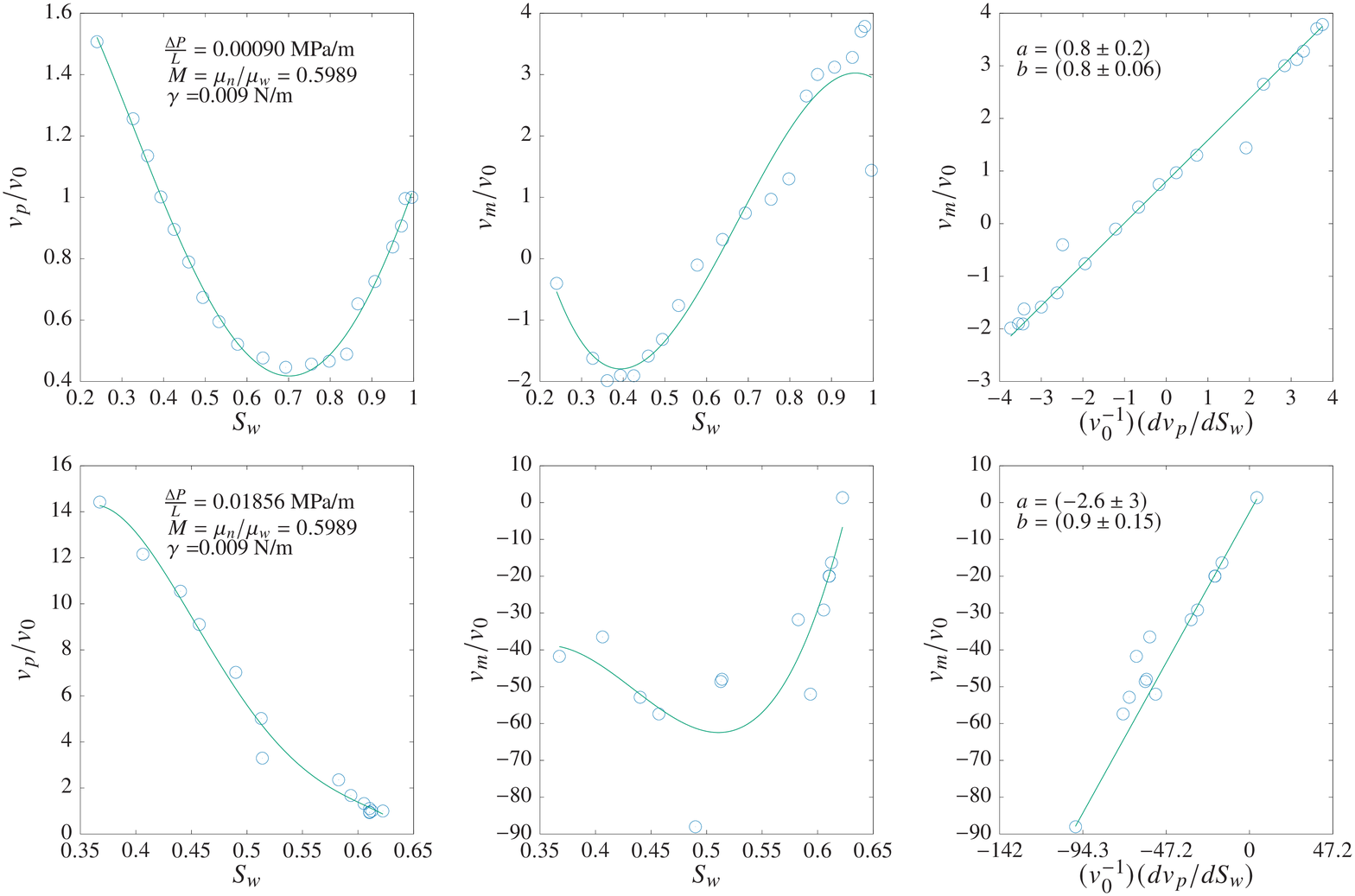} 
\caption{Experimental data from figure 3 in Oak et al.\ \cite{obt90} 
The process is primary drainage (upper row) and imbibition (lower row) 
of a single experimental run. Both rows: Berea sandstone, water as 
non-wetting fluid and oil as wetting fluid, $\phi = 0.193$, 
$K = 1973.8\times10^{-16}$ m$^2$. $\phi$ and $\gamma$ were not supplied 
by the authors.}
\label{fig:Oak_experimental_WO}
\end{figure}

\begin{figure}
\centering
\includegraphics[width=0.8\textwidth,clip]{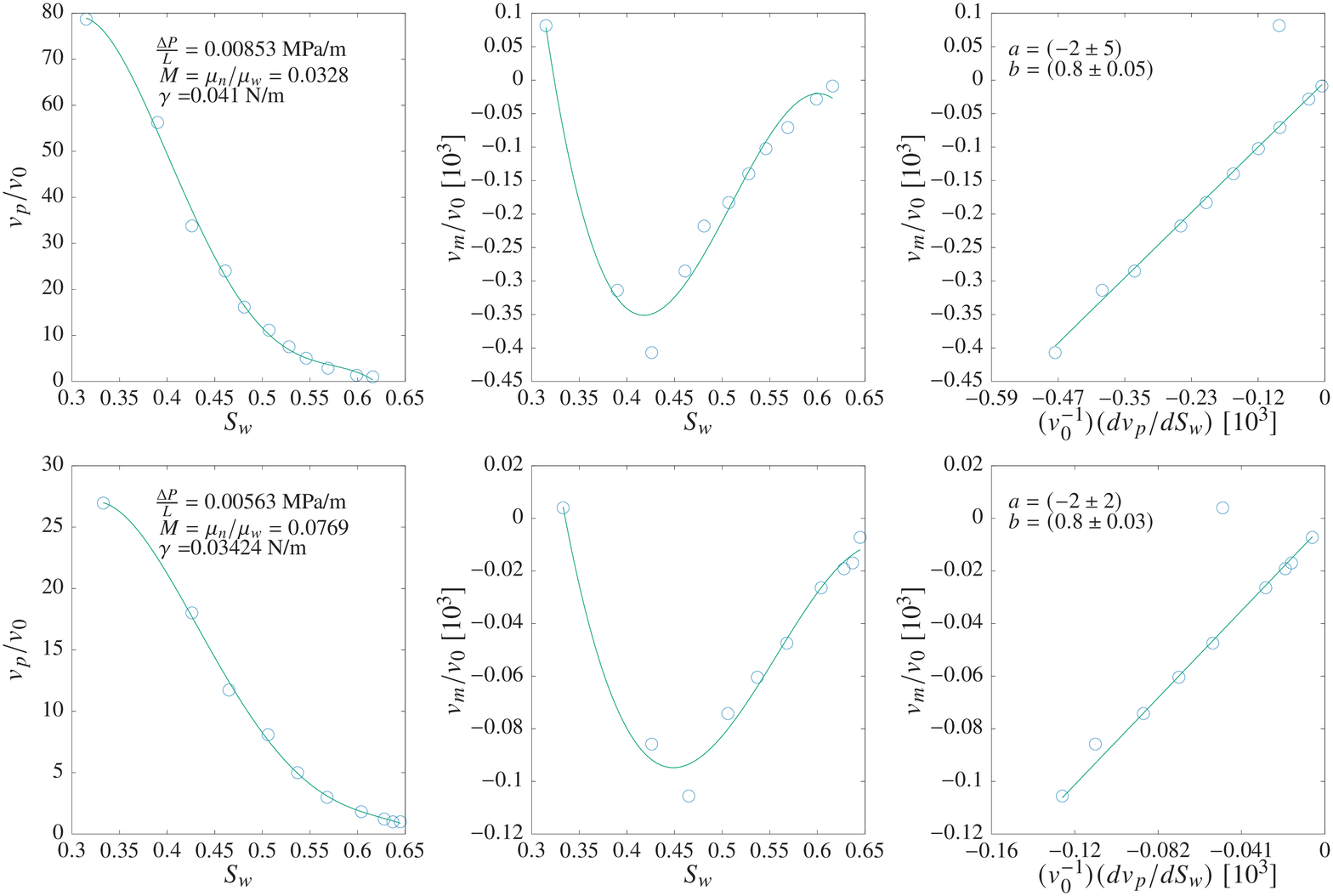}
\caption{Experimental data from runs no.\ $4$ and $5$ in Reynolds et al.\
\cite{rk15}. Upper row: Bentheimer sandstone, drainage with CO$_2$ as 
non-wetting fluid and water as wetting fluid, $\phi = 0.222$, $K = 17862.89\times10^{-16}$ 
m$^2$. Lower row: Bentheimer sandstone, drainage with CO$_2$ as non-wetting fluid and 
brine as wetting fluid, $\phi = 0.222$, $K = 17862.89 \times 10^{-16}$ m$^2$. }
\label{fig:Reynolds_experimental}
\end{figure}

\begin{figure}
\centering
\includegraphics[width=0.8\textwidth,clip]{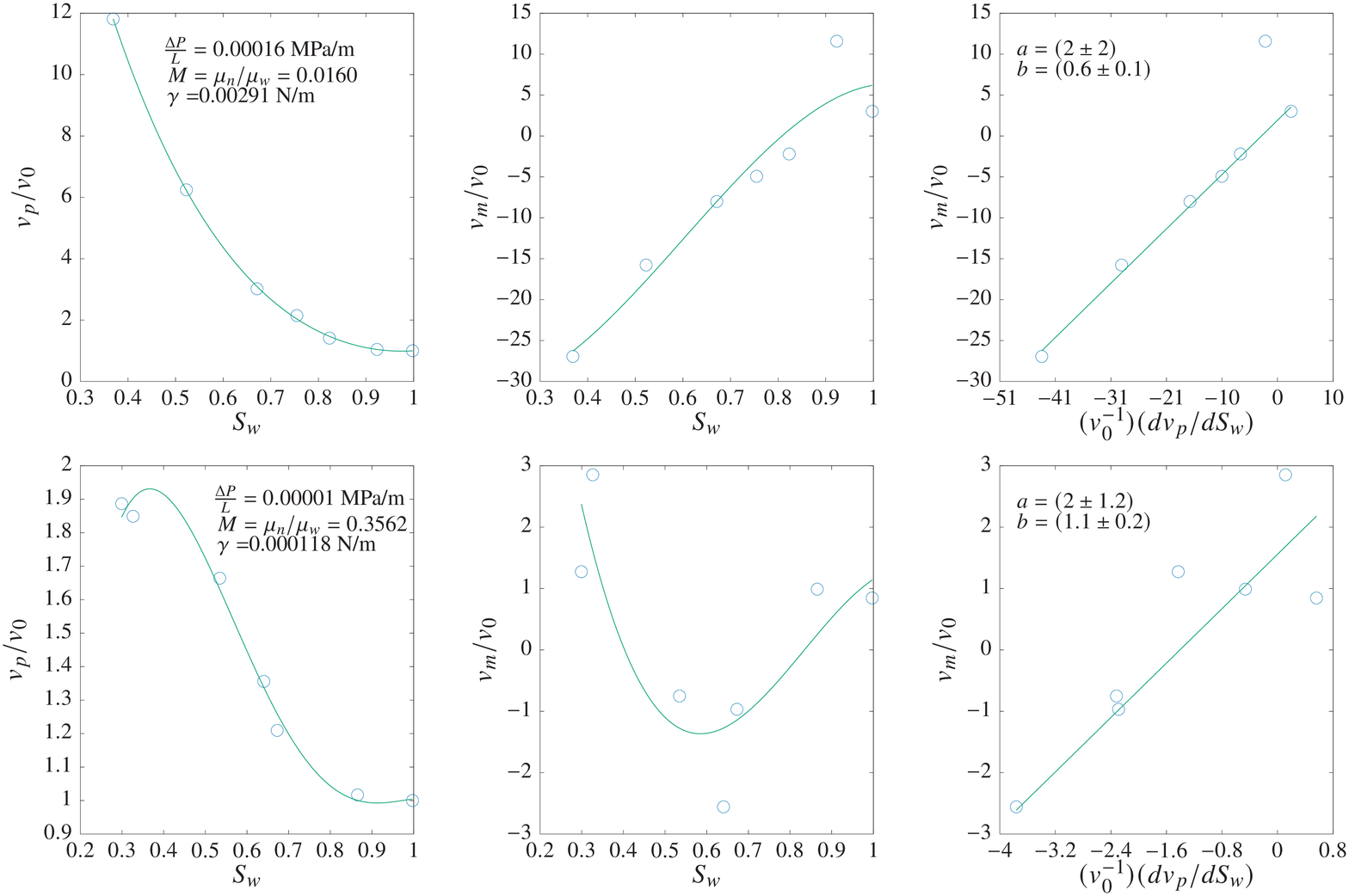} 
\caption{Experimental data extracted graphically using Webplotdigitizer \cite{r20} 
from runs no.\ 18 and 19 in Fulcher et al.\ \cite{fes85}. Both rows: Berea 
sandstone, drainage with oil as non-wetting fluid and water as wetting 
fluid, $\phi = 0.224$. Upper row has $K = 4109.45\times10^{-16}$ m$^2$, 
and lower row $K = 3794.63 \times 10^{-16}$ m$^2$.}
\label{fig:Fulcher_experimental}
\end{figure}

\begin{figure}
\centering
\includegraphics[width=0.8\textwidth,clip]{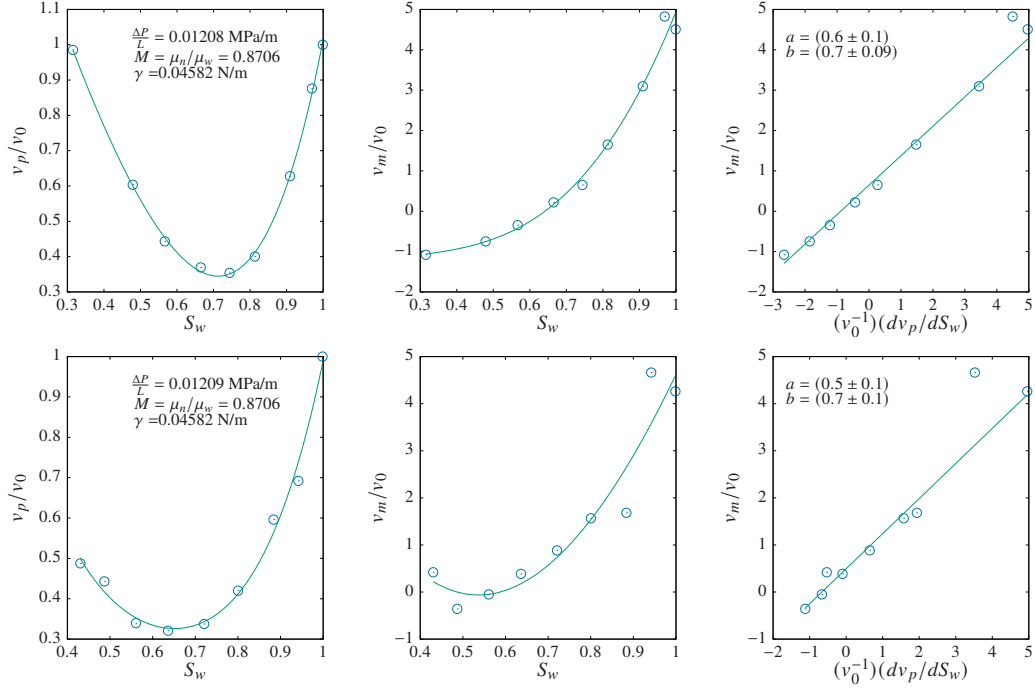} 
\caption{Experimental data extracted graphically from Virnovsky et.al. 
      \cite{vvsl98}, fig. $4$ and $5$. Both rows: Berea sandstone,
      drainage with oil as non-wetting fluid and H$_2$O as wetting fluid, $\phi = 0.561$, $K =
      2131.7\times10^{-16}$ m \textsuperscript{2}.}
    \label{fig:Virnovsky_experimental}
  \end{figure}

\begin{figure}
\centering
\includegraphics[width=0.8\textwidth,clip]{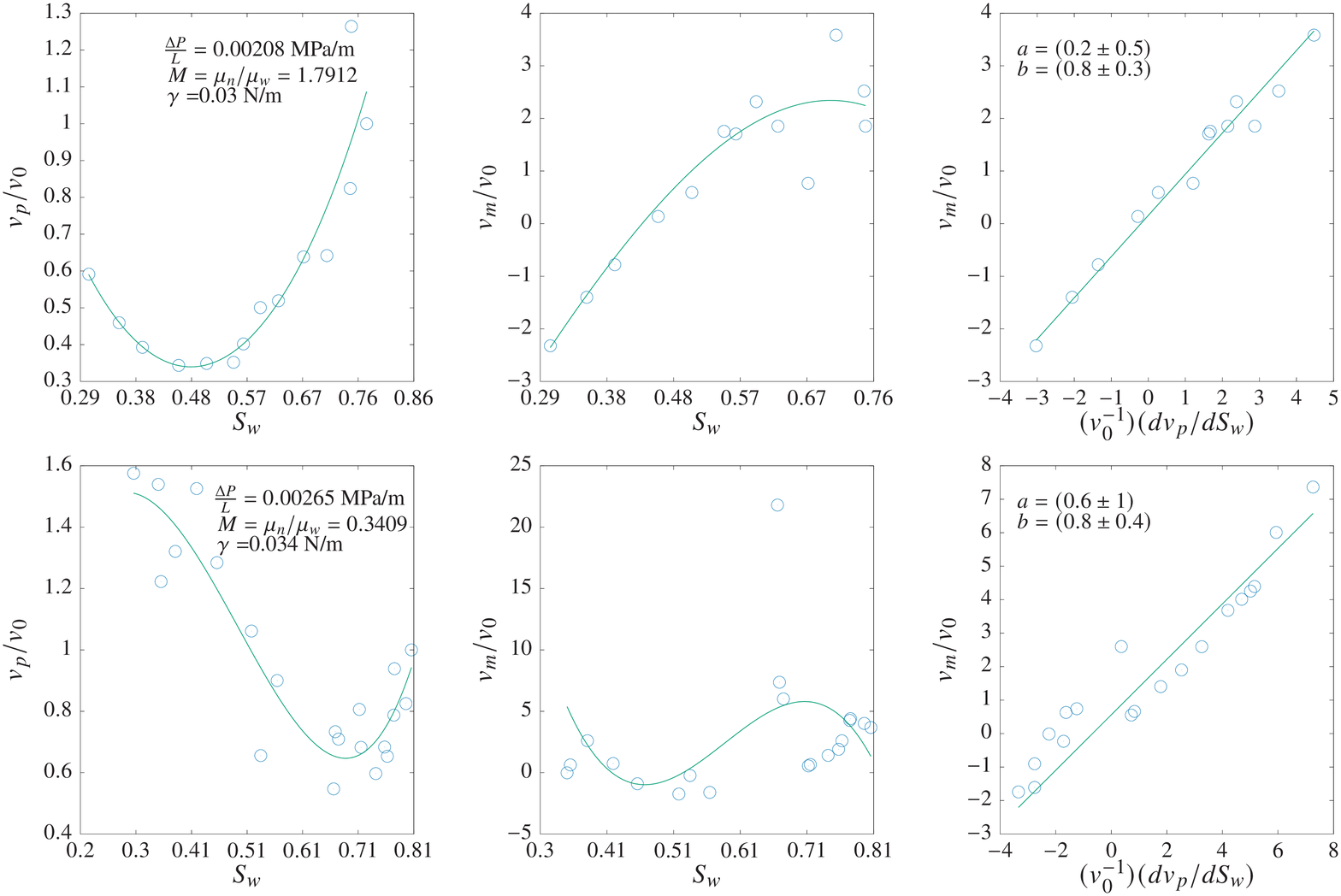} 
\caption{Experimental data extracted graphically using Webplotdigitizer \cite{r20} 
from figure 9 (sand II) in Leverett \cite{l39}. Upper row: drainage in sand, with 
oil as the non-wetting fluid and water as the wetting fluid, $\phi = 0.35$, 
$K = 17270.75\times10^{-16}$ m$^2$. Lower row: sand with oil as the non-wetting fluid
and water as the wetting fluid, $\phi = 0.45$, $K =  10263.76\times 10^{-16}$ m$^2$.}
\label{fig:Leverett_experimental}
\end{figure}

\begin{figure}
\includegraphics[width=0.35\textwidth,clip]{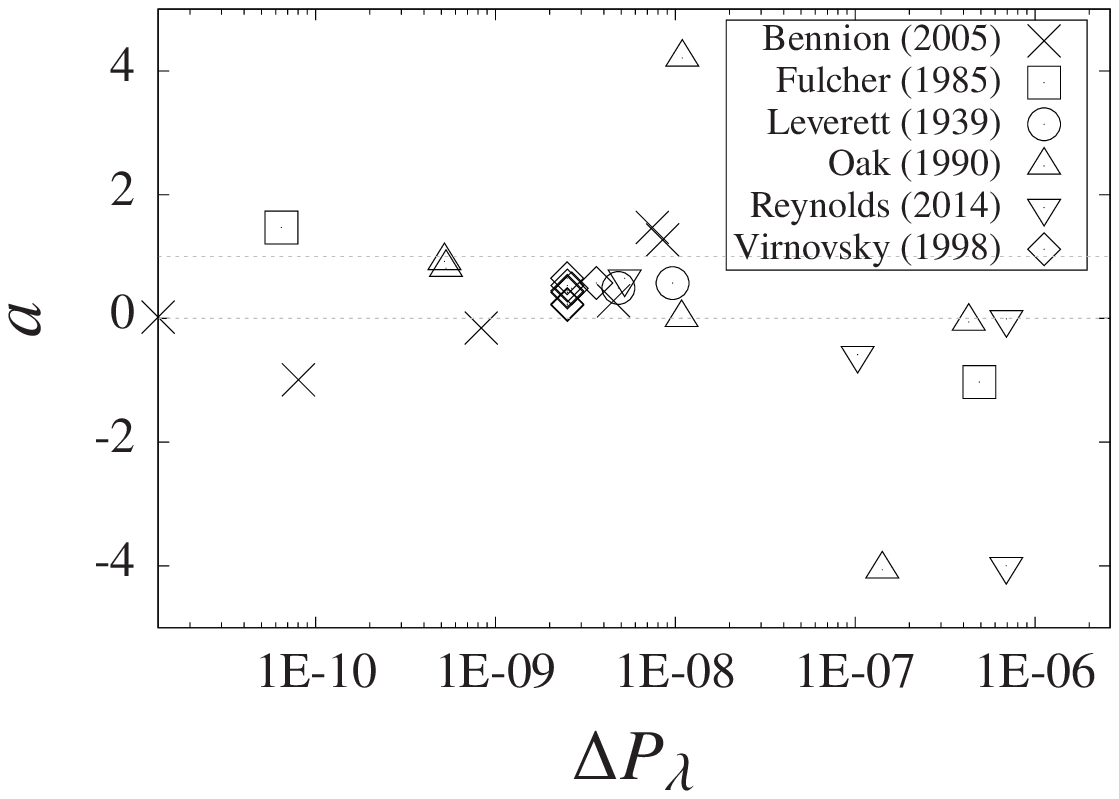} 
\includegraphics[width=0.35\textwidth,clip]{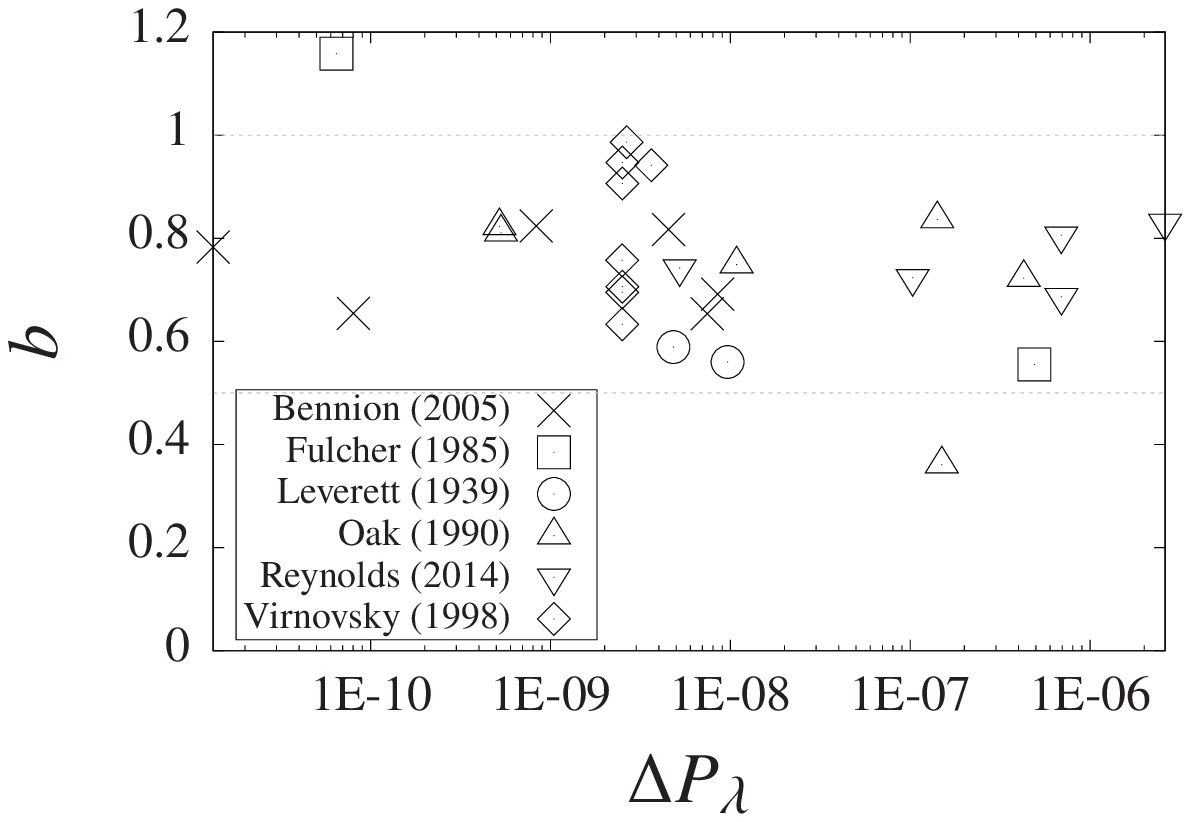} 
\caption{Calculated coefficients $a$ and $b$ from all experimental data as a
function of the scaled pressure gradient $\Delta P_{\lambda}$. In
the left plot, the only data point not shown is $a=-58\pm20$ from the Oak
et al.\ data set \cite{obt90}. In the right plot, the single data point above 
$b=1$ is obtained from a data set with few data points, see figure 
\ref{fig:Fulcher_experimental}.  Values for more data sets than have 
been plotted in figures \ref{fig:Bennion_experimental} --- 
\ref{fig:Leverett_experimental} are included: the rest of the data 
series in Table 3 in Bennion et al.\ \cite{bb05}, an additional water/oil 
data series from figure 3 in Oak et al.\ \cite{obt90}, runs no.\  2, 3 and 6
from Reynolds et al.\ \cite{rk15}, and the two lower flow-rate data 
series from figures 4 and 5 in Virnovsky et al. \cite{vvsl98}.}
\label{fig:ab_coefficients_experimental}
\end{figure}
\subsection{Analysis of relative permeability curves from the literature}
\label{AnalysisRelPerm} 

We analyze in the following relative permeability curves from 
References \cite{bb05,fes85,obt90,vvsl98,rk15,l39} in light of the discussion 
in Section \ref{RelPermEuler}. Our aim is to determine $v_p$ and $v_m$ as a
function of the wetting saturation $S_w$.  

The wetting and non-wetting relative permeabilities $k_{rw}(S_w)$ and
$k_{rn}(S_w)$ data together with the wetting and non-wetting viscosities 
$\mu_w$ and $\mu_{n}$ as supplied by the authors are the essential data
we use in our analysis.  Other parameters such as the surface tension 
$\gamma$, porosity $\phi$ and absolute permeability $K$ we use to set 
the velocity scale $v_0$ and to determine a scale for the pressure gradient.

The data points for $k_{rn}$, $k_{rw}$ and $S_w$ were obtained explicitly from 
tables when available in the cited works. If explicit values were not given, 
the values were extracted graphically from the plots using the software
Webplotdigitizer  \cite{r20}.

In all of the experiments, the measurements were performed when the flow 
reached steady state i.e.\ when the variation in pressure and saturation 
attained values within some acceptable threshold interval. The sources have
different definitions of when steady state has been reached, but this 
threshold is usually taken to be fluctuations within $1 - 2 \% $ 
over the span of minutes to hours depending on the experiment, see \cite{obt90}. 

The data we use were obtained either during drainage or imbibition
processes. 

We plot the velocities $v_p$ in equation (\ref{RelPereq04a}) and $v_m$ in
equation (\ref{RelPereq05}) in dimensionless units by dividing
by a velocity scale $v_0$ and $\mu_w$. We do this in the following way:
We define
\begin{equation}
\label{RelPereq04c}
\tilde{v}_p(S_w)=\left[\frac{k_{rw}(S_w)}{\mu_w}+\frac{k_{rn}(S_w)}{\mu_n}\right]\;,
\end{equation}
and
\begin{align}\label{RelPereq05f}  
\tilde{v}_m (S_w)= \left[\frac{S_w}{\mu_w}\frac{d}{dS_w}
\left(\frac{k_{rw}(S_w)}{S_w}\right)+\frac{S_n}{\mu_n}
\frac{d}{dS_w}\left(\frac{k_{rn}(S_w)}{S_n}\right)\right]\;.
\end{align}
We then define
\begin{equation}
\label{RelPereq04f}
\tilde{v}_0=\tilde{v}_p(S_w=1)\;,
\end{equation}
leading to
\begin{equation}
\label{RelPereq04d}
\frac{v_p(S_w)}{v_0}=\frac{\tilde{v}_p(S_w)}{\tilde{v}_0}\;,
\end{equation}   
and
\begin{equation}
\label{RelPereq05d}
\frac{v_m(S_w)}{v_0}=\frac{\tilde{v}_m(S_w)}{\tilde{v}_0}\;.
\end{equation}
It is the right hand side of these two equations that we plot. 

The first and second columns of figures where we present our data analysis 
\ref{fig:Bennion_experimental} --- \ref{fig:Leverett_experimental} show
plots of $v_p(S_w)/v_0$ and $v_m(S_w)/v_0$ while the third column shows $v_m/v_0$ plotted
against $d(v_p/v_0)/dS_w$.  We have used both equations (\ref{RelPereq05}) and
(\ref{relperm-1}) to determine $v_m$.  They are of course in principle 
equivalent, but they demand different numerical differentiations. Both 
gave the same result. It is the values of $v_m$ calculated from equation 
(\ref{relperm-1}) that are shown in the plots. 

The experimental data shown in the plots in this section have not been
picked based on any special criteria. However, we have prioritized data 
sets with a larger number of data points for the plots.  

We now turn to the results of our analysis.  The third column of figures
\ref{fig:Bennion_experimental} --- \ref{fig:Leverett_experimental}
shows $v_m/v_0$ as a function of $v'_p/v_0$ where $v'_p=dv_p/dS_w$. The 
surprising result is that the relation 
\begin{equation}
\label{relperm-200}
v_m=av_0+b\frac{dv_p}{dS_w}\;,
\end{equation}
where $a$ and $b$ are constants with respect to $v'_p$, fits the data 
excellently.  {\it This is our main result.\/}

The parameters $a$ and $b$ in (\ref{relperm-200}) have been determined by
finding the visually best straight line for each data set. These best lines 
are shown in the figures. The quality of the fits vary. The data in 
figure \ref{fig:Bennion_experimental} fit the best to a straight line, whereas 
the data that fit to a straight line the least are found in figure  
\ref{fig:Fulcher_experimental}.  This is reflected in the uncertainty of 
the coefficients $a$ and $b$. The uncertainty is in general larger in $a$
than in $b$.  Note that the $a$ and $b$ coefficients for drainage and 
imbibition in figure \ref{fig:Oak_experimental_WG} and 
\ref{fig:Oak_experimental_WO} are slightly different. 

In the first two columns of figures 
\ref{fig:Bennion_experimental} --- \ref{fig:Leverett_experimental}
where we have plotted $v_p$ and $v_m$ against $S_w$, we have fitted 
the data to polynomials; for $v_p$ we have used fourth order polynomials 
for the numerical fits and for $v_m$ we have used third order 
polynomials. The reason for this lies in equation (\ref{relperm-200}) 
which indicates that $v_m$ should be modeled with a polynomial of 
one less order than that of $v_p$. The $v_p$-polynomial is
numerically fitted directly to the data. For the $v_m$ fit, one can either I: 
fit a third order polynomial directly to the data, or II: calculate the coefficients
for the $v_m$ fit using those found for $v_p$ using equation (\ref{relperm-200}).
In principle, these two methods should give the same results. However, method
II is highly sensitive to noise in the data series. Method II was used for all 
the data set, and the correspondence is good between the fit and
the data for the sets with the lowest amount of deviation, figure
\ref{fig:Bennion_experimental} in particular. Here, method II showed only small
deviations from method I in the initial and final values of the data series.
Method I was used in all of the plots, as the method of fit for $v_p$ and $v_m$
does not affect the rest of the results.

We plot in figure \ref{fig:ab_coefficients_experimental} the values of the
coefficients $a$ and $b$ as a function of the pressure gradient for all the 
data series. The pressure $\Delta P$ is rendered dimensionless by dividing it
by $\mu_w v_0/K$, $\Delta P_\lambda=\Delta P K/\mu_w v_0$. 

\section{The average seepage velocity $v_p$ and the co-Moving velocity $v_m$ 
in a dynamic pore network model}
\label{poreNetwork}

The porous medium is represented by a network of nodes and links
in dynamic pore network modeling \cite{jh12}. The immiscible fluids 
are transported through the links which are connected at the nodes.
The dynamic pore network model we consider here was introduced by 
Aker et al.\ \cite{amhb98}. A recent review describe it in detail, 
see \cite{sgvh20}. 

The nodes do not contain fluid, only the links do. The nodes only 
represent the points where the links meet. The flow rate $q_j$ inside any 
link $j$ of the network at any instant of time is obtained by \cite{shbk13,w21},
\begin{equation}
  \displaystyle
  q_j = -\frac{g_j}{l_j\mu_{av}}\left[\Delta p_j - \sum p_{c,j}\right]\;,
  \label{eqnWB}
\end{equation}
where $l_j$ is the link length, $g_j$ is the link permeability which depends on the 
cross section of the link and $\Delta p_j$ is the pressure drop across link. 
The viscosity term $\mu_{av}$ is the saturation-weighted viscosity of the 
fluids inside the link given by $\mu_{av} = s_{j,w}\mu_w + s_{j,n}\mu_n$ where 
$s_{j,w}$ and $s_{j,n}$ are the wetting and non-wetting fluid saturations inside the link. 
The term $\sum p_{c,j}$ is the total interfacial pressure 
from the fluid interfaces in the link $j$. A pore typically 
consists of two wider pore bodies connected by a narrow pore throat. We model this 
by using hour-glass shaped links. The variation of the interfacial pressure with the 
interface position for such a link is modeled by \cite{shbk13}
\begin{equation}
  \displaystyle
  |p_c\left(x\right)| = \frac{2\gamma\cos\theta}{r_j}
  \left[1-\cos\left(\frac{2\pi x}{l_j}\right)\right]\;,
  \label{eqnpc}
\end{equation}
where $r_j$ is the average radius of the link and $x \in [0,l_j]$ 
is the position of the interface inside the link. Here, $\theta$ is the 
contact angle between the interface and the pore wall and $\gamma$ is 
the surface tension between the fluids. 

These two equations (\ref{eqnpc}) and (\ref{eqnWB}), together with the Kirchhoff relations, 
i.e., the sum of the net volume flux at every node at each time step will be 
zero, provide a set of linear equations. In order to calculate the local flow rates, 
we solve these equations with a conjugate gradient solver \cite{bh88}. All the interfaces 
are then advanced accordingly using small time steps. 

In order to achieve steady-state flow, we apply 
periodic boundary conditions in the direction of flow.

We use a two-dimensional square lattice with $64 \times 64$ links with link 
lengths $l_j = 1\,{\rm mm}$. Disorder is introduced by 
choosing the link radii $r_j$ randomly from a uniform distribution 
in the range $0.1\,{\rm mm}$ to $0.4\,{\rm mm}$. We use $100$ different 
realizations of such networks for our simulations. 

Assuming Poiseuille flow in the links, the average link permeability 
$r_j^2/8$ will be $7.8\times 10^{-7}$ m$^2$.  As it is a square lattice,
the length of it compensates for its width, making its permeability
equal to the link permeability times $\sqrt{2}$ to account for its 
$45^\circ$ tilt. This gives an estimate for the average permeability
of the lattice around $5.5\times 10^{-7}$ m$^2$.          

We will in the following explore $v_p$ and $v_m$ as a function of the wetting 
saturation $S_w$ defined as the total volume of wetting fluid in the 
links divided by their total pore volume, and the average pressure gradient
defined as $\Delta P/L$ where $\Delta P$ is the pressure difference across the 
network. The viscosity ratio $M$ is defined as the ratio  of the viscosity 
$\mu_n$ of non-wetting fluid to the viscosity $\mu_w$ of the wetting fluid 
($M=\mu_n/\mu_w$).

\subsection{Fitting the average seepage velocity $v_p$ and the co-moving velocity $v_m$
to polynomials}
\label{vp_sw}

We discuss here $v_p$ and $v_m$ as a function of the wetting
saturation $S_w$ for fixed pressure gradient $\Delta P/L$.  The
average seepage velocity $v_p$ is measured directly from the model.  The co-moving
velocity is inferred from the velocity difference $v_n-v_w$ and the derivative
$dv_p/dS_w$ according to equation (\ref{eqn19}).  
We fit the data to the polynomials 
\begin{equation}
\label{vpnpoly}
v_p=\sum_{k=0}^4 C_k S_w^k\;,
\end{equation}
and  
\begin{equation}
\label{vmnpoly}
v_m=\sum_{k=0}^3 D_k S_w^k\;.
\end{equation}
We find that the three or fourth order polynomials form an adequate compromise between accuracy and 
the wish to keep the number of fitting parameters down.  

\begin{figure}[ht]
\centerline{\includegraphics[width=0.7\textwidth,clip]{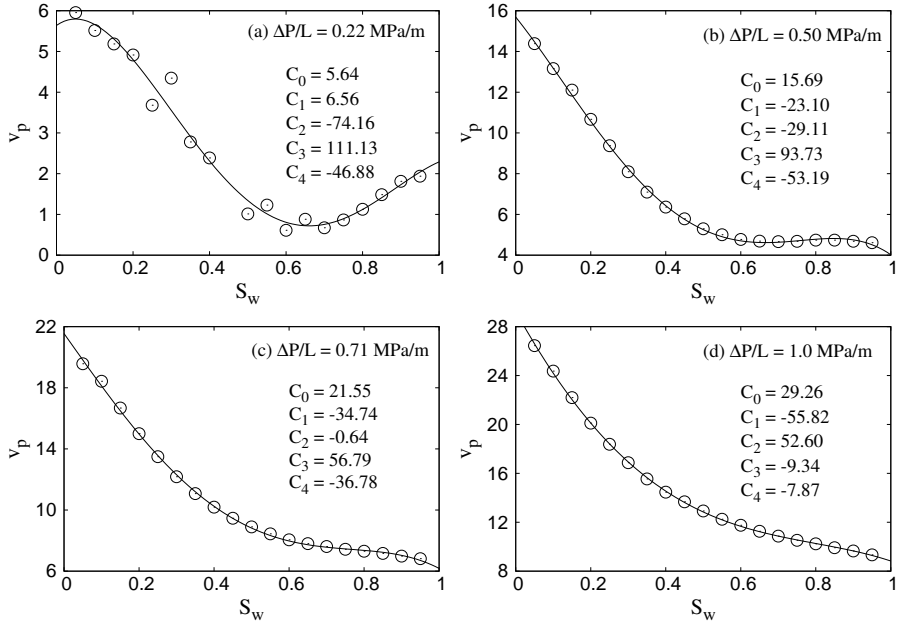}}
\caption{Fitting of the numerical results for $v_p$ vs $S_w$ with equation (\ref{vpnpoly}) 
with $\mu_w=0.03$ Pa s and $\mu_n=0.01$ Pa s (i.e., $M=1/3$), and for four different pressure 
gradients, $\Delta P/L=$ 0.22, 0.50, 0.71 and 1.0 MPa/m.  The fitting parameters are shown in the 
legends in each figure. The rate of change of $v_p$ ($v_p^{\prime}=dv_p/dS_w$) 
will be calculated from this figure and will be used to express $v_m$ in terms of 
$v_p^{\prime}$ in figure \ref{fig3}.}
\label{fig2}
\end{figure}

Figure \ref{fig2} shows how the seepage velocity $v_p$ behaves as a function of the 
wetting saturation $S_w$ for four different pressure gradients: 
$\Delta P/L$ = 0.22, 0.5, 0.71 and 1.0 MPa/m. 
The results are obtained for $0.05 \le S_w \le 0.95$ with intervals of 0.05, totaling 
20 data points. For now, we use $\mu_w=0.03$ Pa s and $\mu_n=0.01$ Pa s, i.e.,
$M=\mu_n/\mu_w=1/3$. The effect of a varying viscosity ratio will be explored 
later in this paper. We observe the quality of the fits to improve with increasing 
pressure gradient.   

\begin{figure}[ht]
\centerline{\includegraphics[width=0.7\textwidth,clip]{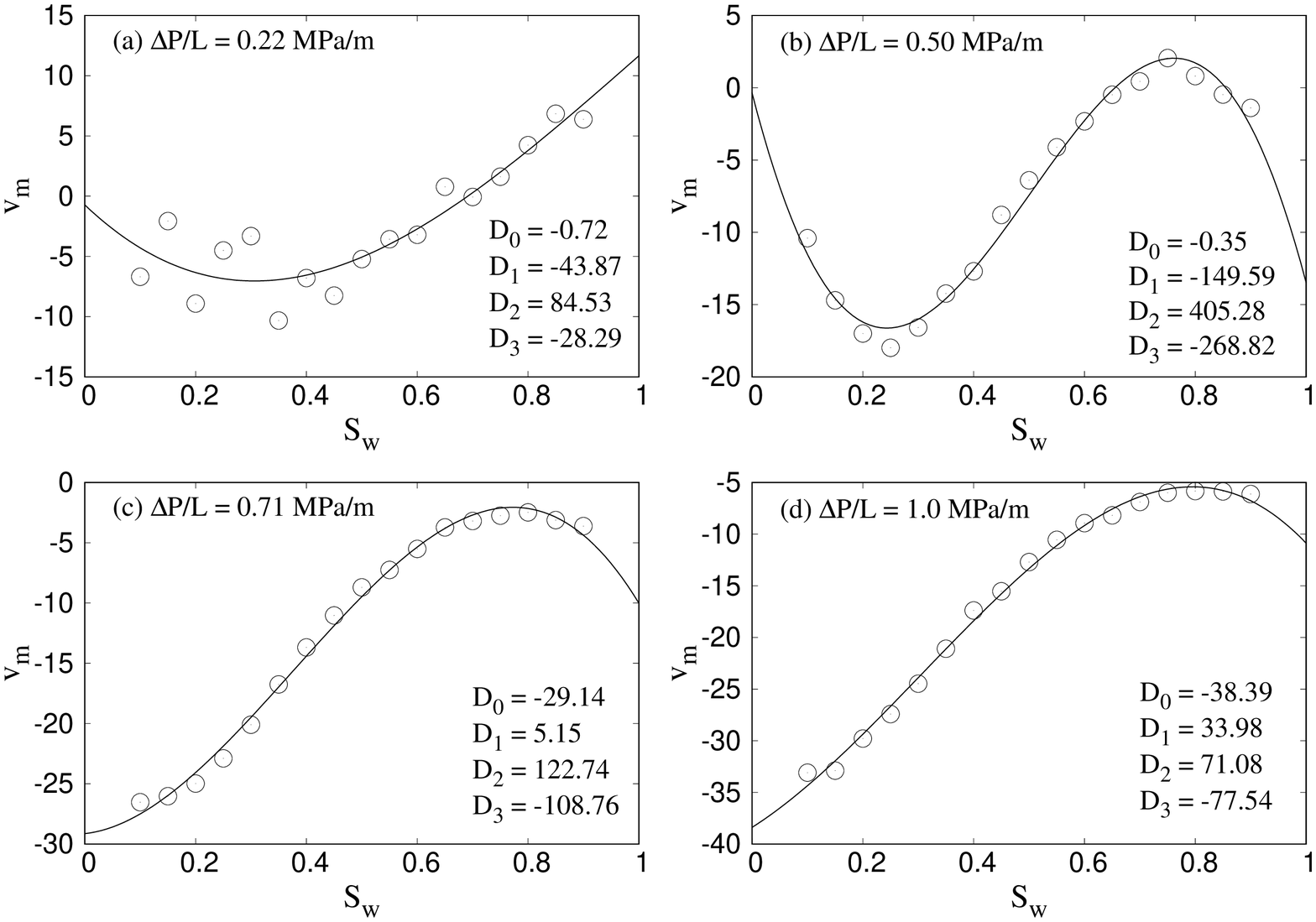}}
\caption{Fitting of the numerical results for $v_m$ vs.\ $S_w$ using equation (\ref{vmnpoly}). 
The parameters $M$ and $\Delta P/L$ are as in Figure \ref{fig2}.}
\label{fig3}
\end{figure}

We use the data in figure \ref{fig2} to approximate $v_p^{\prime}$ by central differencing,
which then is used to determine $v_m$ from equation (\ref{eqn19}).  
We show the result in figures \ref{fig2} where we plot $v_m$ as a function of $S_w$.       

\begin{figure*}[t]
\includegraphics[width=0.4\textwidth,clip]{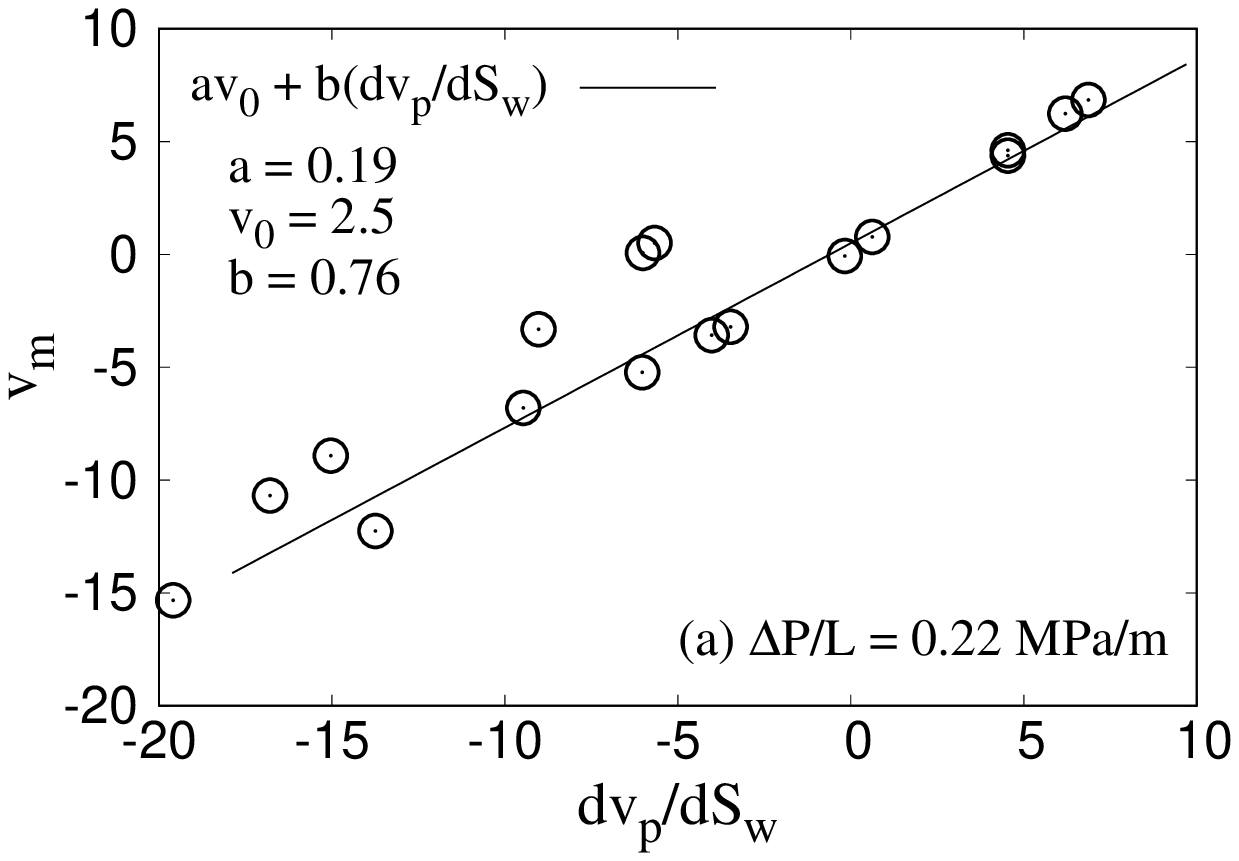} 
\includegraphics[width=0.4\textwidth,clip]{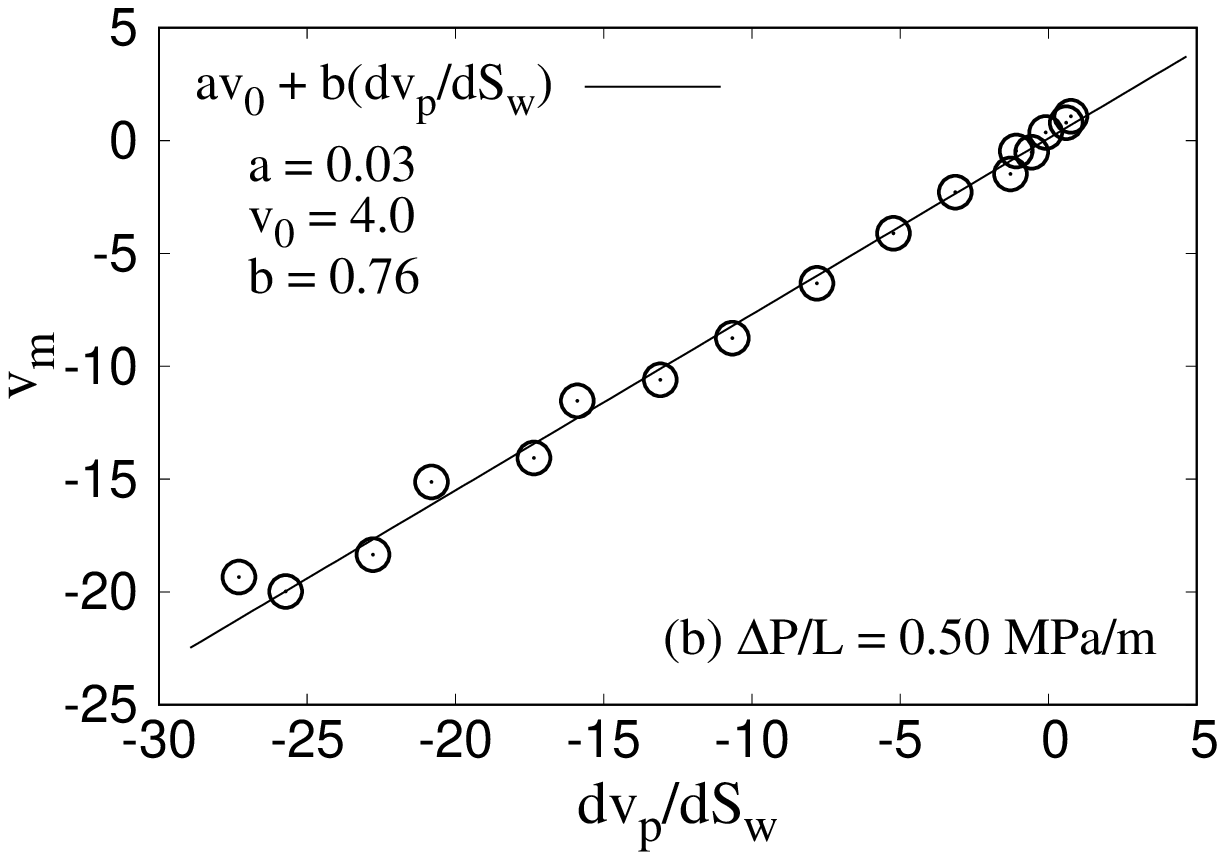} \\
\includegraphics[width=0.4\textwidth,clip]{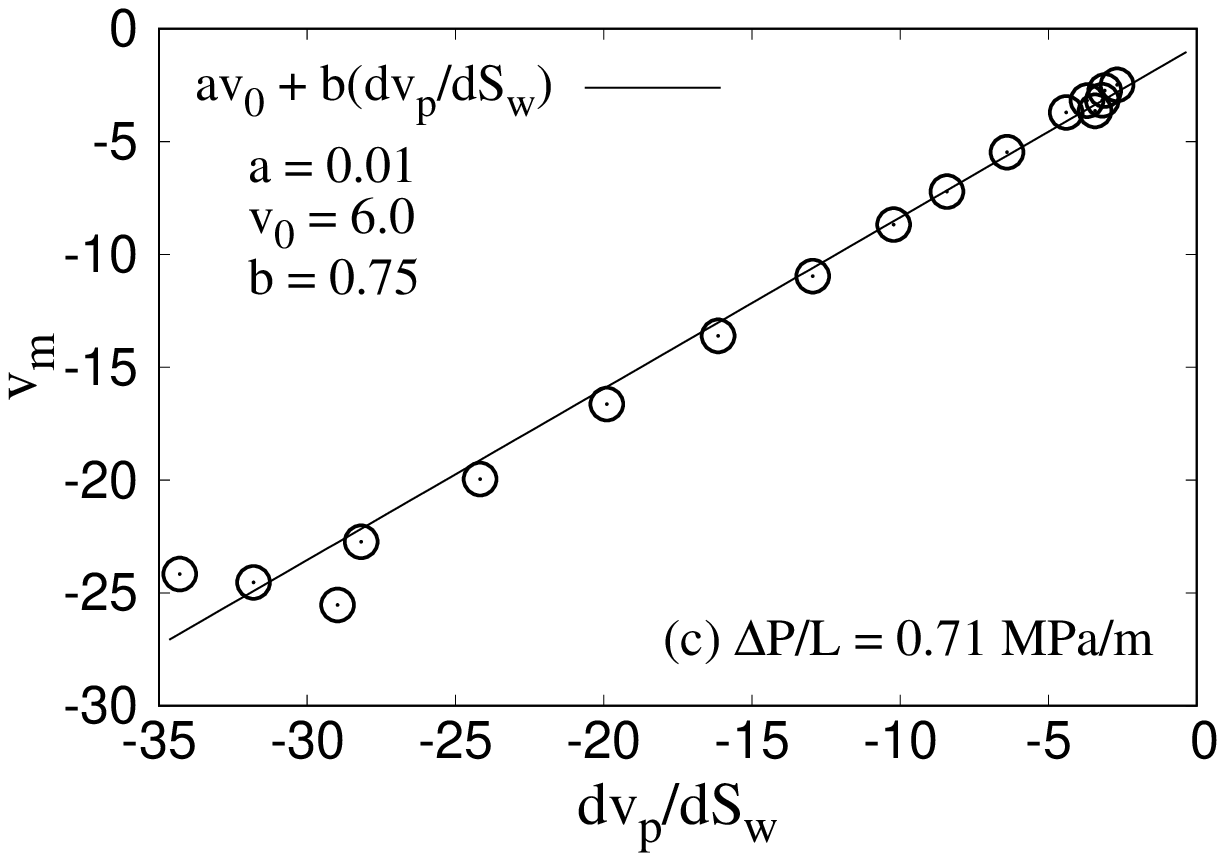} 
\includegraphics[width=0.4\textwidth,clip]{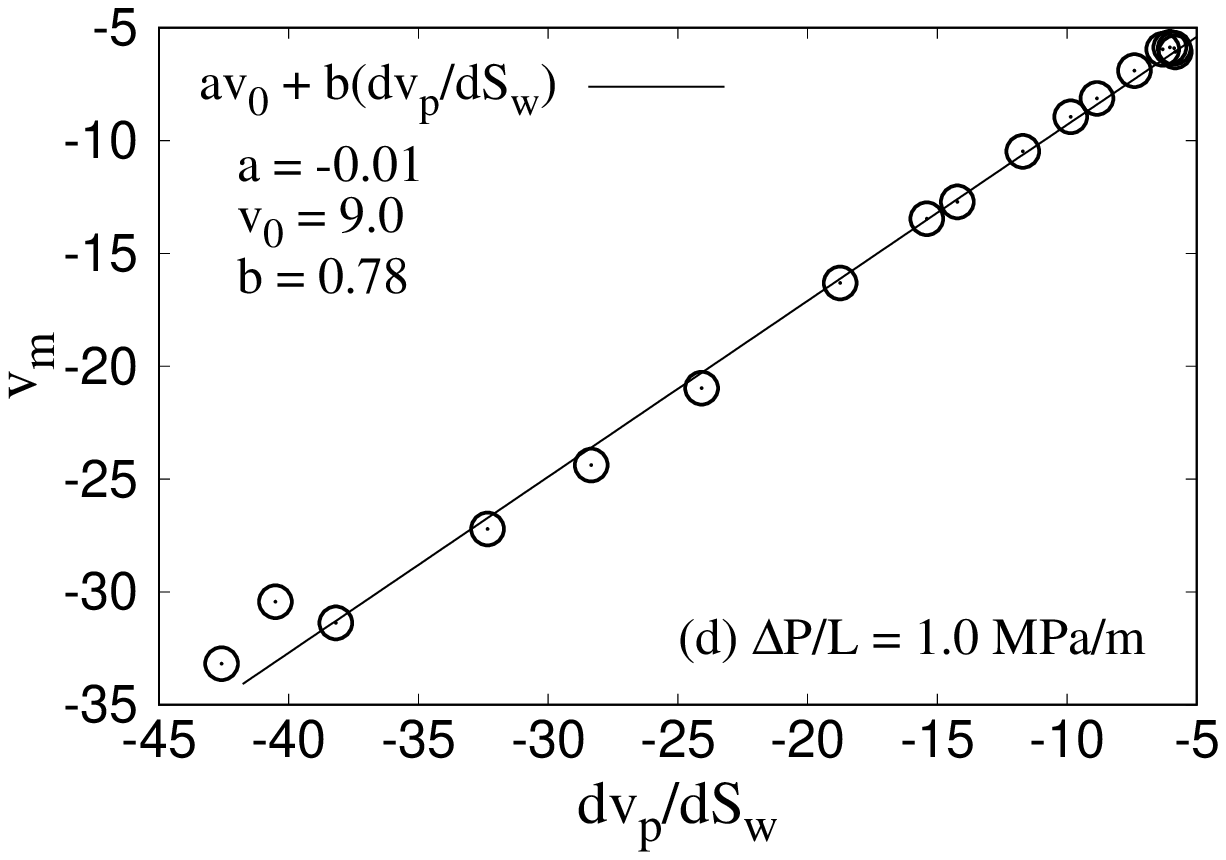} \\
\includegraphics[width=0.4\textwidth,clip]{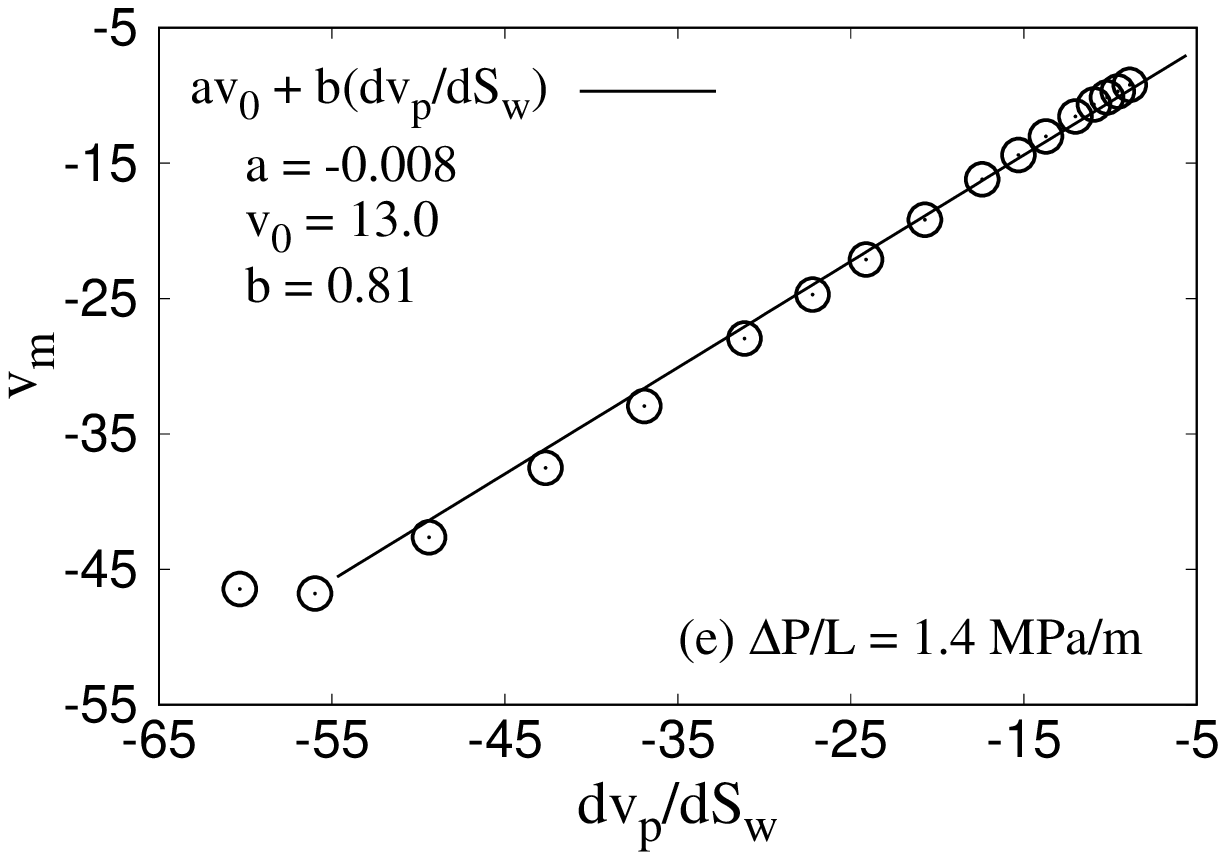} 
\includegraphics[width=0.4\textwidth,clip]{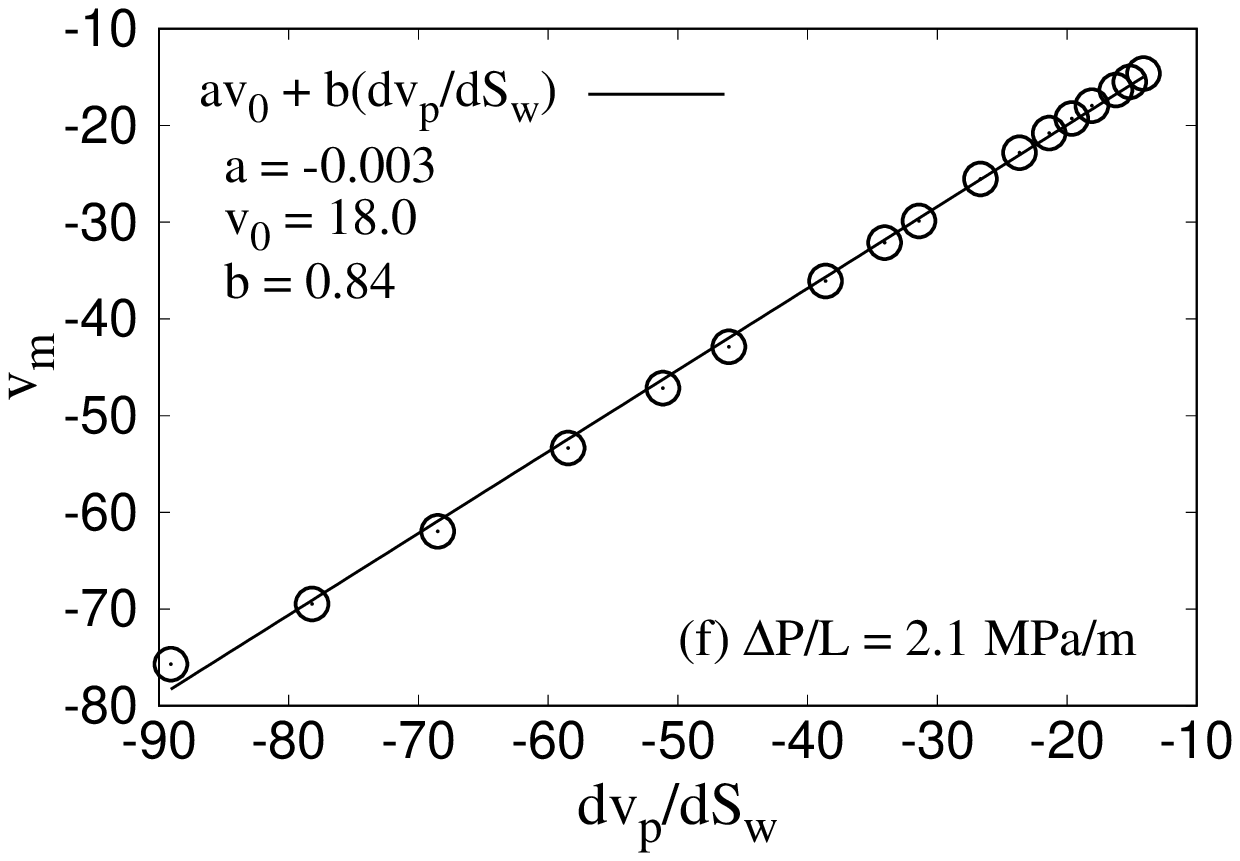}
\caption{The co-moving velocity as a function of $dv_p/dS_w$ from equation (\ref{eqn19}) 
for $\mu_w=0.03$ Pa s and $\mu_n=0.01$ Pa s, and $\Delta P/L=$ 0.22, 0.5, 0.71, 1.0, 
1.4 and 2.1 MPa/m.}
\label{fig4}
\end{figure*}

We plot $v_m$ against $v_p^{\prime}$ in figure \ref{fig4} for fixed $\Delta P/L=$ 0.22,
0.50, 0.71, 1.0, 1.4 and 2.1 MPa/m.  We introduce a velocity scale $v_0=v_p(S_w=1,\Delta P/L)$
to make the fits comparable to the relative permeability-based fits we discussed in Section 
\ref{expResults}.  As is evident, equation (\ref{relperm-200}) fits the data well. 
We note that both $a$ and $b$ vary with the pressure gradient $\Delta P/L$.  Hence, we 
write equation (\ref{relperm-200}) as 
\begin{equation}
\label{eq07}
v_m\left(S_w,\frac{\Delta P}{L}\right)=a\left(\frac{\Delta P}{L}\right)
v_0\left(\frac{\Delta P}{L}\right)+
b\left(\frac{\Delta P}{L}\right)\ \frac{dv_p}{dS_w}\left(S_w,\frac{\Delta P}{L}\right)\;.
\end{equation}
We have in this equation written explicitly what parameters each variable depends upon. This will
become important in the next section.   

\subsection{The co-moving velocity when $dv_p/dS_w$ is treated as an independent variable}
\label{older_vm}

The co-moving velocity $v_m$ has been calculated using the dynamic network model in both
References \cite{hsbkgv18} and \cite{sgvh20}.  In contrast to our approach here, the
derivative $v_p^{\prime}$ was treated as an independent variable in those papers. 
That is, $v_m$ was plotted against $(S_w,v_p^{\prime})$ producing a plane.  
In \cite{hsbkgv18}, a variant of the dynamic
pore network model we use here was used \cite{gvkh19}, resulting in the relation
\begin{equation}
\label{old_vm_1}
v_m\left(S_w,\frac{dv_p}{dS_w}\right)=c+d\ S_w+e\ \frac{dv_p}{dS_w}\;,
\end{equation}
where $c\approx -0.095$, $d\approx -0.15$ and $e\approx 0.79$ for data averaged 
over both square and hexagonal lattices. Sinha et al.\ considered both a square lattice
and a lattice based on a reconstructed Berea sandstone, giving $c=5.00\pm 0.13$, $d=-6.36\pm 0.25$ 
and $e=0.94\pm 0.01$ for the square lattice and $c=10.10\pm 0.32$, $d=-12.94\pm 0.62$ 
and $e=0.88\pm 0.01$ for the reconstructed Berea sandstone. 

Equation (\ref{eq07}) constitutes a cut through the plane $(S_w,v_p^{\prime})$ given
by $\Delta P/L$ constant.  It is an open question as to why the explicit $S_w$ dependence
dependence disappears in equation (\ref{eq07}) when making this cut.        

\subsection{Dependence of coefficients $a$ and $b$ on the pressure gradient}
\label{a_b_deltaP}

\begin{figure}[ht]
\includegraphics[width=0.4\textwidth,clip]{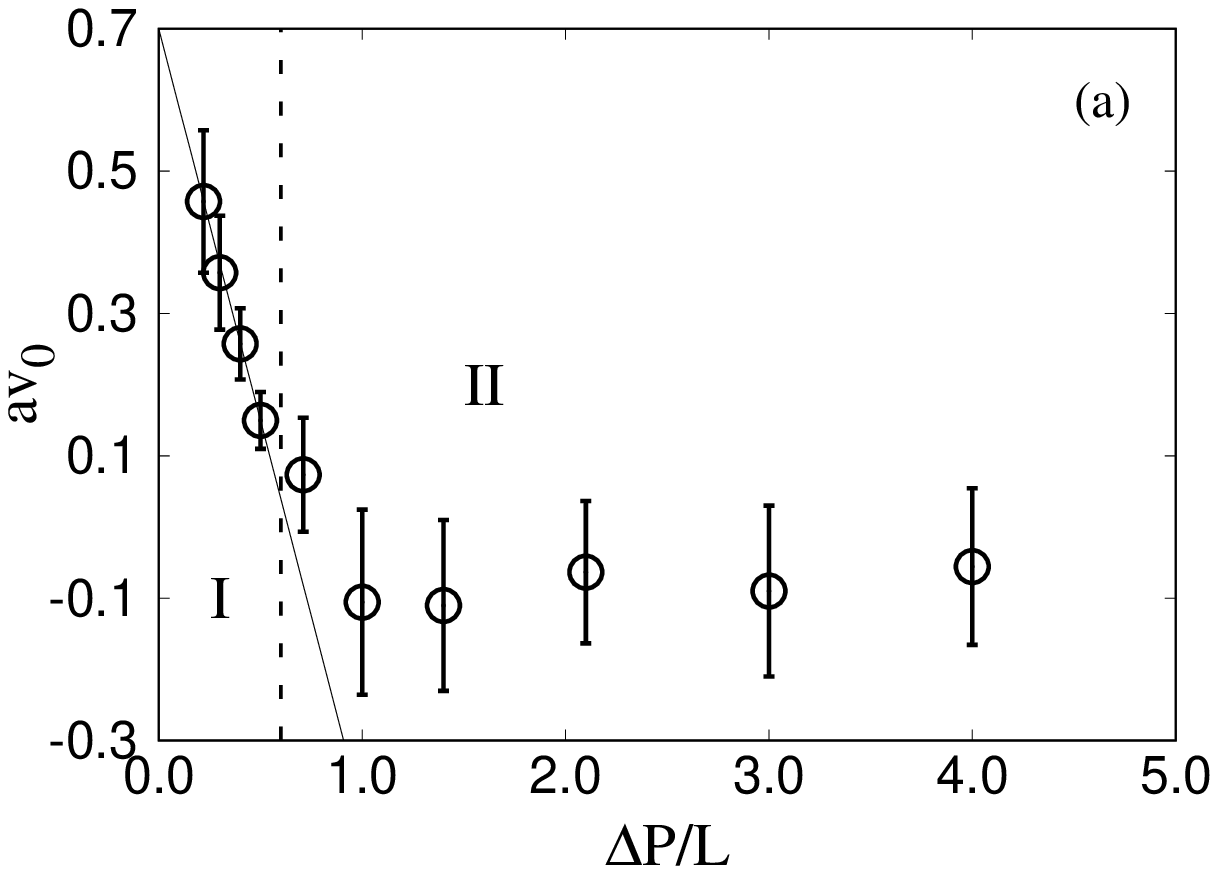} 
\includegraphics[width=0.4\textwidth,clip]{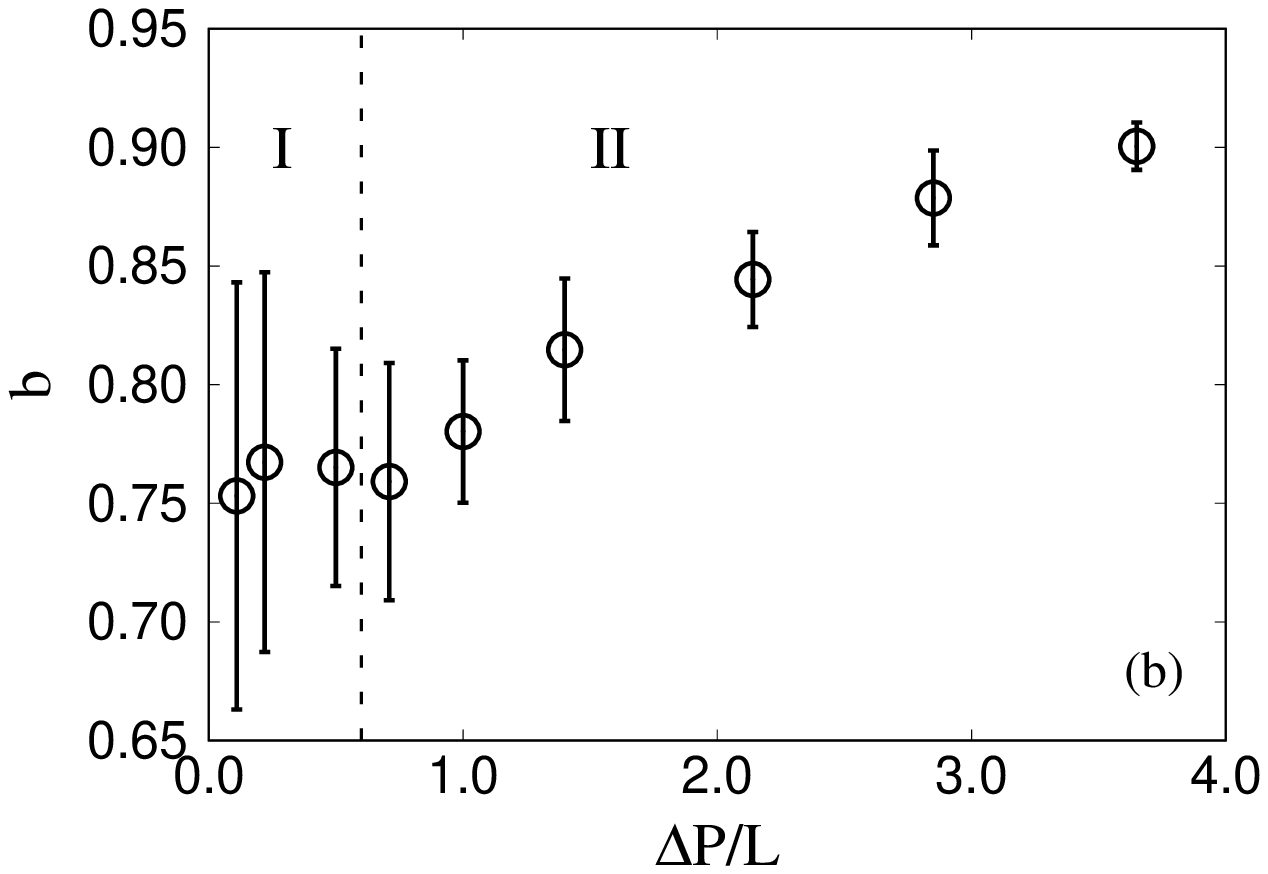}
\caption{(a) and (b) respectively shows $av_0$ and $b$ defined in equation (\ref{eq07}) 
as a function of the pressure gradient $\Delta P/L$. The viscosities of the fluids were
$\mu_w=0.03$ Pa s and $\mu_n=0.01$ Pa s, so that $M=1/3$. We have marked two regions, 
I and II in both (a) and (b). The straight line in region I in (a) is 
$0.7m/s-1.1(\Delta P/L)m^2/MPa s$.}  
\label{fig5}
\end{figure}

Figure \ref{fig5} shows the variation of  $av_0$ and $b$ defined in equation (\ref{eq07}), 
as a function of the pressure gradient $\Delta P/L$. 
We observe two different regions as the fluid velocities increase with increasing 
pressure gradient. We name these regions I and II.\\  

{\it Region I ---\/} This is the low pressure gradient region. We find a good fit to
the data with the line $av_0=0.7m/s-1.1(\Delta P/L)m^2/MPa s$. The coefficient $b$ has a value
around 0.76.  Due to low flow velocity, the $av_0$ and $b$ found in this region can be 
compared with the relative permeability data in Section \ref{expResults}.\\     

{\it Region II ---\/} This is the high pressure gradient region. Here $av_0$ saturates to 
a value near $-0.1$ m/s whereas $b$ approaches the value 1 asymptotically. 
This is outside the region where the relative permeability data would be relevant.\\  

The crossover of $av_0$ from positive to negative value and the onset of increment 
in $b$ is observed to take place around the same pressure gradient.

\subsection{$v_p$ as a function of $S_w$ and $\Delta P/L$}
\label{vp_nonlinear}

We now turn to the average seepage velocity $v_p$.  As described in the Introduction, there is
a regime over an interval of pressure gradients where the flow rate is proportional
to the pressure gradient to a power  \cite{tkrlmtf09,tlkrfm09,aetfhm14,sh17,glbb20,zbglb21}.
This regime is clearly visible in our dynamic pore network model \cite{sh12,fsrh21}. 
Our aim in this section is to map out the $v_p$ over a wide range of saturations $S_w$ 
and pressure gradients $\Delta P/L$.  

\begin{figure}[ht]
\centerline{\includegraphics[width=0.6\textwidth,clip]{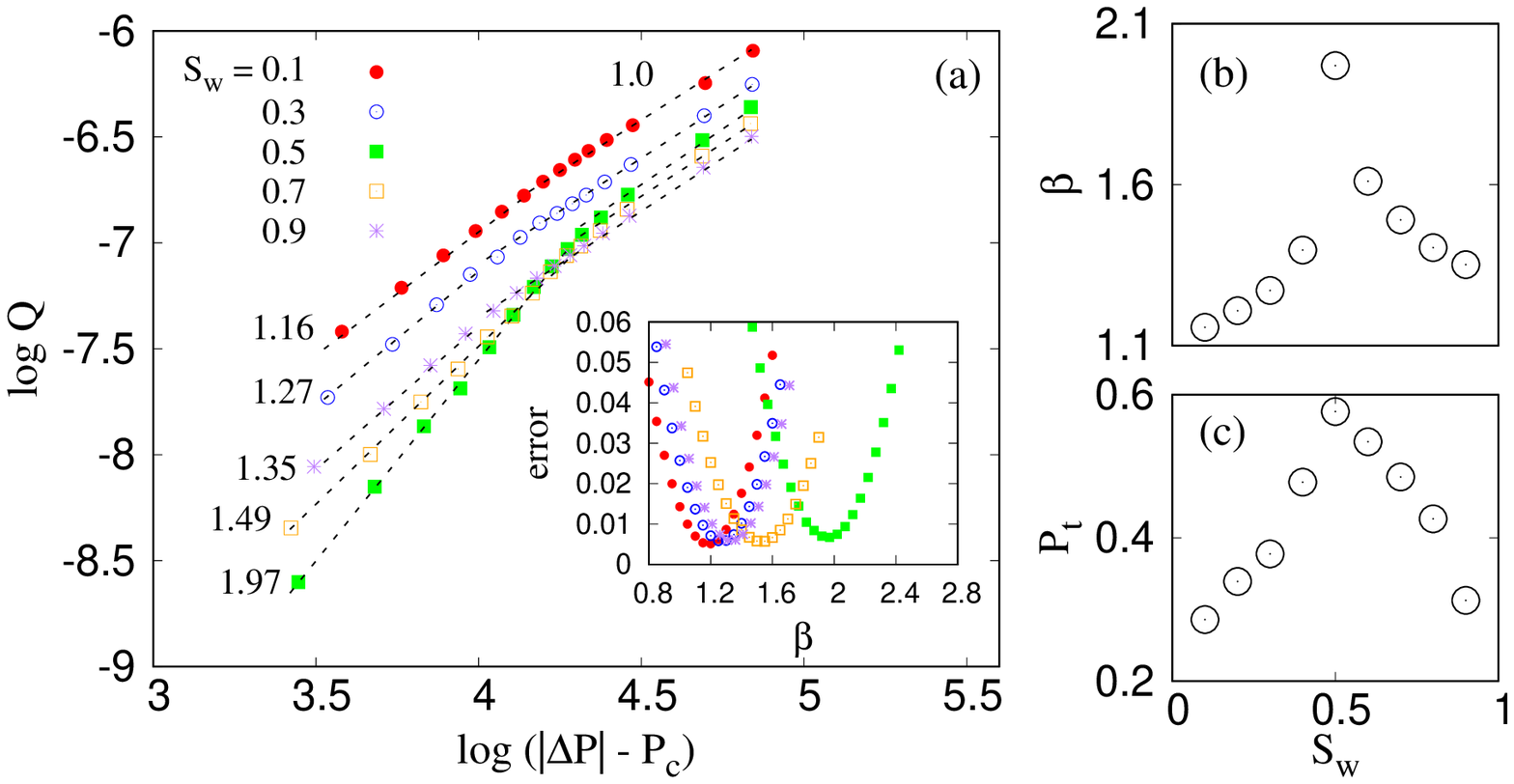}}
\caption{(a) shows $Q$ vs.\ $\Delta P$  for wetting saturations $S_w=$ 0.1, 0.3, 0.5, 0.7 and 0.9. 
The viscosities of the fluids were $\mu_w=0.03$ Pa s and $\mu_n=0.01$ Pa s. The inset shows 
the fitting error between the data and equation (\ref{eq08}) as a function of $\beta$ for the
different wetting saturations. (b) and (c) show the dependence of $P_t$ and $\beta$ on $S_w$.}
\label{fig6}
\end{figure}

Figure \ref{fig6}(a) shows how the flow rate $Q$ increases as the pressure gradient 
$\Delta P/L$ increases. We observe the following behavior:

\begin{equation}
\label{eq08}
Q\propto \left\{\begin{array}{ll}
0       & \mbox{, $|\Delta P| \le P_s$\;,}\\
\left(\left|\frac{\Delta P}{L}\right| -\frac{P_s}{L}\right)^{\beta} & \mbox{, $|\Delta P| > P_s$\:,}\\
              \end{array}
       \right.
\end{equation}
where $P_s$ is a threshold pressure below which there is no flow. This threshold is 
a finite-size effect, see \cite{rsh19b}. Above a pressure difference $|\Delta P|\gg P_t$,
the exponent $\beta=1$ and we observe Darcy-like linear flow. Below this pressure
difference, $P_s < |\Delta P| < P_t$, the exponent $\beta > 1$.    

The inset in figure \ref{fig6}(a) demonstrates how the threshold pressure $P_s$ and $\beta$ were
calculated: For a constant $S_w$, we first set a particular $\beta$ value and fit the numerical 
results to equation (\ref{eq08}), finding $P_s$ as well as the error associated with the fit. 
In this way we get a $P_s$ value and an error value as a function of $\beta$. The curves in the 
inset show the error as a function of $\beta$ for different saturations. We identify the minimum
of the error vs.\ $\beta$ curve.  The value of $\beta$ giving the error minimum and the corresponding 
$P_s$ value are the values we assign to the system for that saturation $S_w$. 

Figures \ref{fig6}(b) and (c) show the variation of exponent $\beta$ and the transition point $P_t$ 
as functions of the wetting saturation $S_w$. Both $\beta$ and $P_t$ is observed to have a maximum at 
$S_w=0.5$ to decrease on both sides of it.  We have that $\beta=1$ for $S_w=0$ and $S_w=1$ as we 
are then dealing with single fluid flow. 

\begin{figure}[ht]
\centerline{\includegraphics[width=0.8\textwidth,clip]{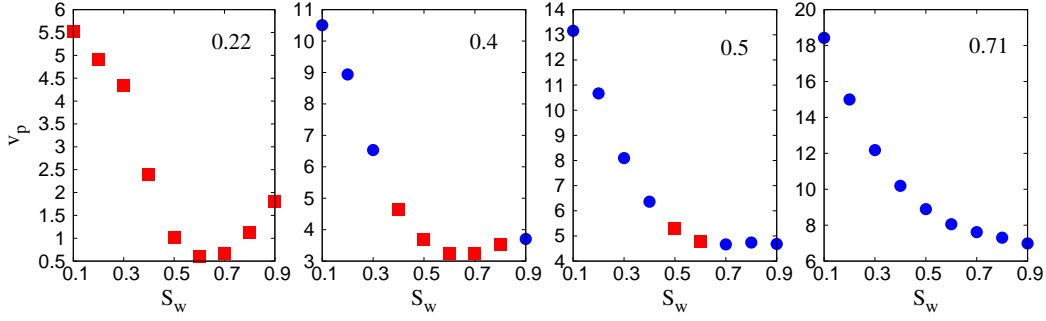}}
\caption{$v_p$ as a function of saturation $S_w$ for $\Delta P/L=0.22$, 0.40, 0.50 and 0.71 
MPa/m respectively. The viscosities of the fluids were $\mu_w=0.03$ Pa s and $\mu_n=0.01$ Pa s. 
A red square indicates that the flow is in the non-linear region for the set of parameters, 
$\Delta P/L$ and $S_w$, that produce this data point. A blue circle indicates that the flow 
is in the linear region.}
\label{fig7}
\end{figure}
\begin{figure}[ht]
\includegraphics[width=0.4\textwidth,clip]{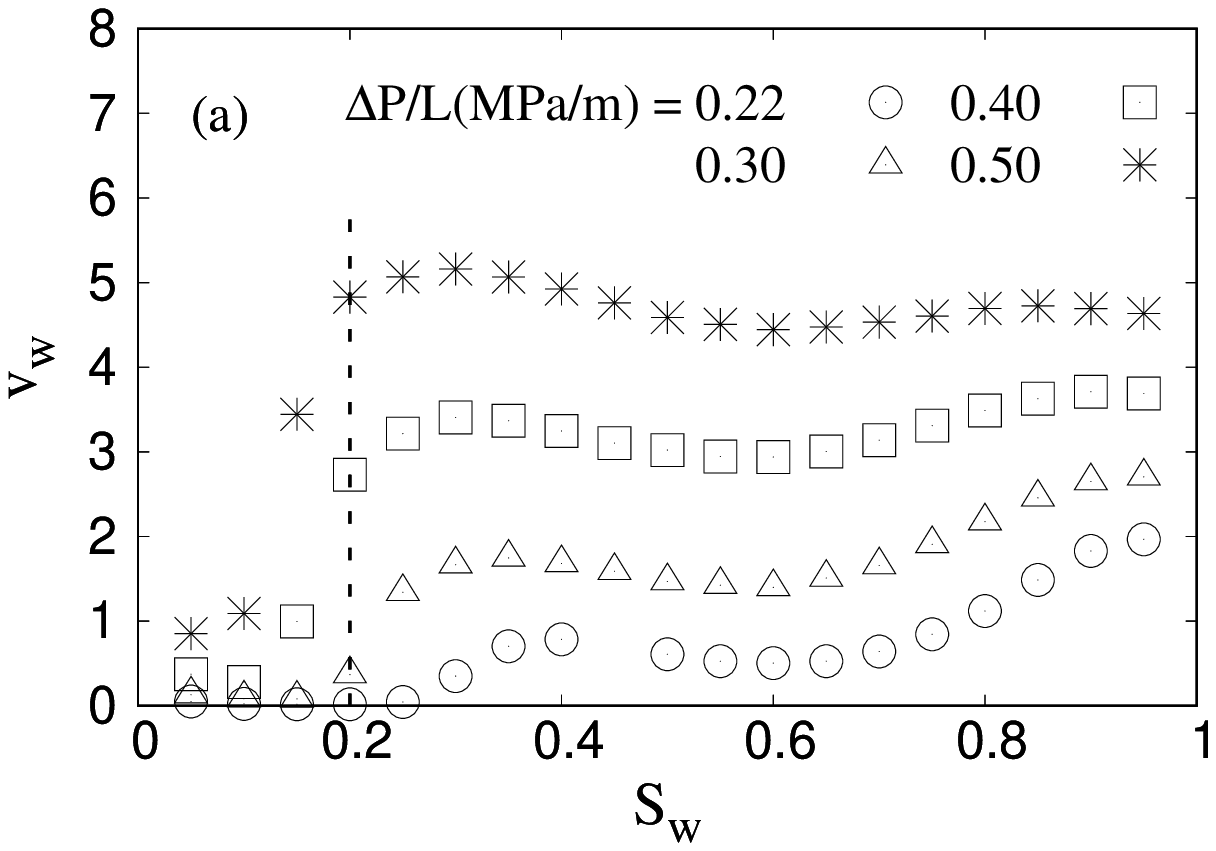} 
\includegraphics[width=0.4\textwidth,clip]{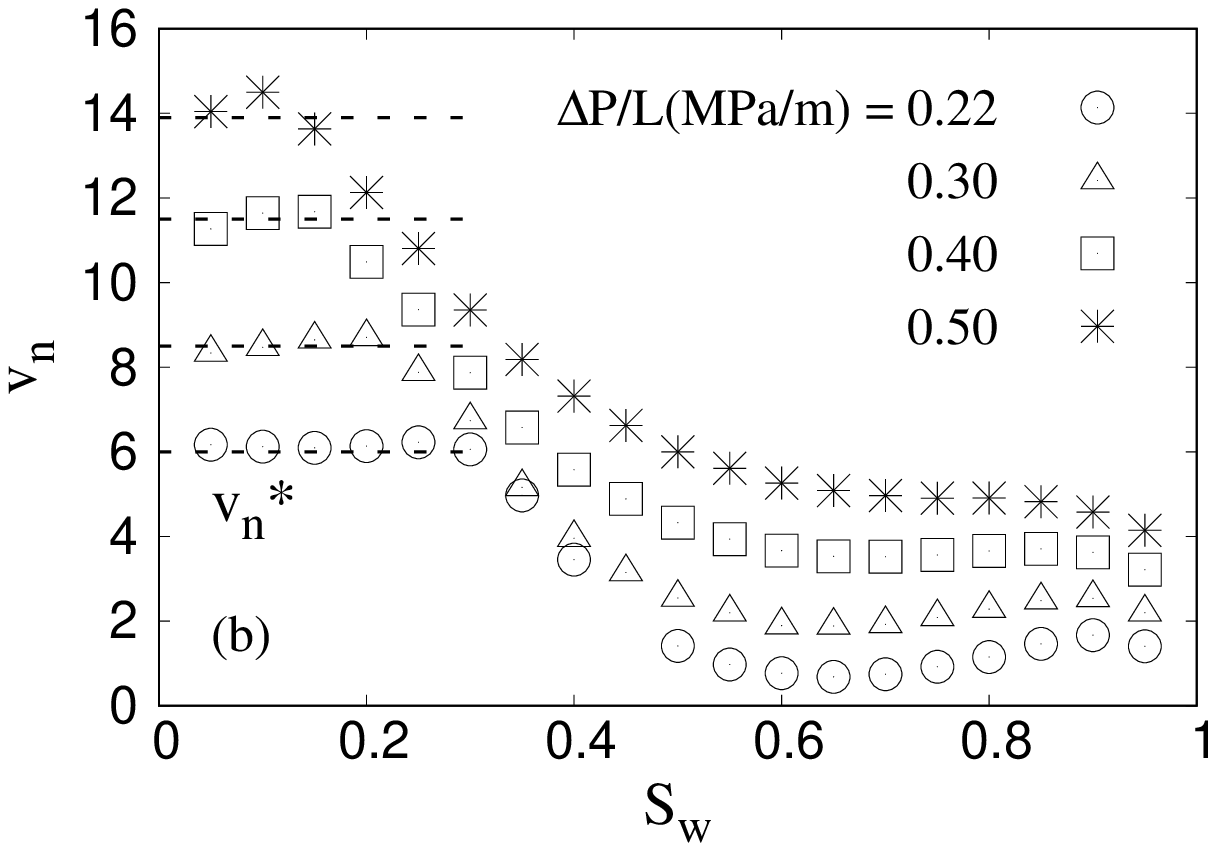} \\ 
\includegraphics[width=0.4\textwidth,clip]{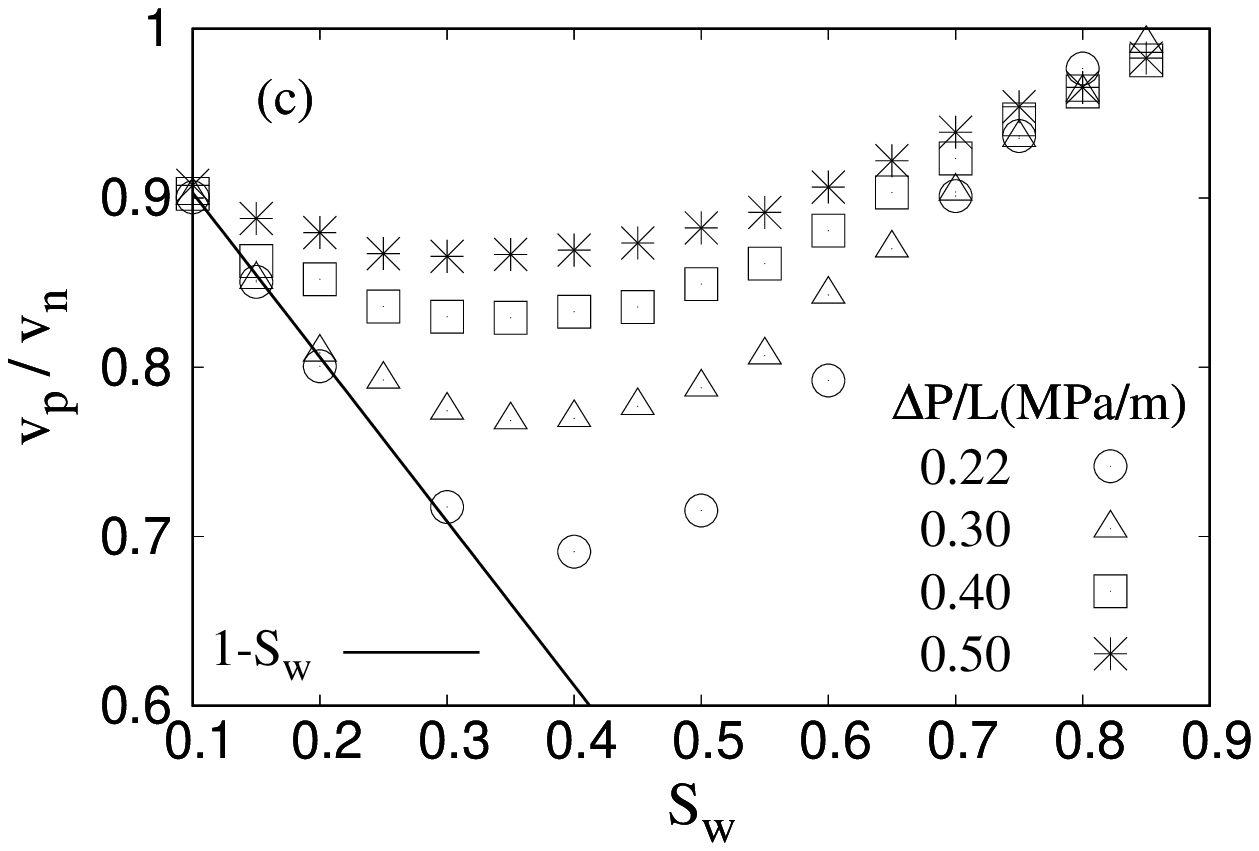} 
\caption{(a) $v_w$ a function of the wetting saturation $S_w$ for $\Delta P=0.22, 0.30, 0.40$ 
and $0.50$ MPa/m. The vertical dotted line shows the value of $S_{w,\min}(\Delta P)$ below which 
$v_w \approx 0$. (b) $v_n$ as a function of the wetting saturation $S_w$. The horizontal dotted 
lines show $v_n^*$ when $S_w<S_{w,\min}$. (c) shows the comparison of the numerical results 
with equation (\ref{new_eq03}) (see the dotted line) for all 4 pressure gradients. 
The viscosities of the fluids were $\mu_w=0.03$ Pa s and $\mu_n=0.01$ Pa s.}
\label{fig10}
\end{figure}

Figure \ref{fig7} shows the flow rate $Q=v_p A_p$ (equation (\ref{eqn5.4})) as a function of 
the wetting  saturation $S_w$ for four different pressure gradients, $\Delta P/L=0.22$, 0.40, 0.50 
and 0.71 MPa/m.  The data points shown as red squares indicate that the flow is in the non-linear
regime where $\beta>1$, i.e., $P_s < |\Delta P| < P_t$.  The data points shown as blue circles 
indicate that the flow is in the linear regime, i.e., $|\Delta P| > P_t$. Hence, we see that for a
range of pressure gradients, e.g., $\Delta P/L=0.4$ MPa/m, $v_p$ visits both the linear and 
non-linear regimes over the range of wetting saturations $S_w$. For pressure gradients larger
than 0.5 MPa/m, $v_p$ is always in the linear regime over the entire range of $S_w$. 

We now compare figures \ref{fig6} and \ref{fig7}.  We note that the transition between the 
linear and non-linear regimes in figure \ref{fig7} appears at essentially the same pressure
gradient that separates regimes I and II in figure \ref{fig6}.  This points towards a 
connection.  However, such a connection is yet to be found.      

\subsection{Limits}
\label{limits}

The irreducible wetting saturation $S_{w,irr}$ is the minimum
wetting saturation possible irrespective of the pressure gradient.
The residual non-wetting saturation $S_{n,r}$
is the minimum non-wetting saturation possible irrespective 
of the pressure gradient. At any finite pressure gradient 
$\Delta P/L$ there will be a minimum wetting saturation 
$S_{w,\min}(\Delta P/L)$ which approaches $S_{w,irr}$ as
the pressure gradient is increased.  Likewise, there will
be for any finite pressure gradient a minimum non-wetting
saturation $S_{n,\min}(\Delta P/L)$ which approaches 
$S_{n,r}$ as the pressure gradient in increased. Let us 
define $S_{w,\max}(\Delta P/L)=1-S_{n,\min}(\Delta P/L)$.
When $S_w$ reaches $S_{w,\min}(\Delta P/L)$
or $S_{w,\max}(\Delta P/L)$, either the wetting or the non-wetting
fluid stops moving.      

Knudsen and Hansen \cite{kh06} demonstrated that there is hysteresis 
at $S_{w,\min}(\Delta P/L)$ based on a dynamic pore network model
closely related to the one we use here, see their Figure 2. The way
Knudsen and Hansen did this was to increase or decrease the saturation step 
by step, building on the steady-state configurations that already were 
established at the previous saturation.

In the numerical work we present here based on the dynamic pore network
model, we re-initiate the model every time we we change the saturation.
This means that the system for each value of the saturation chooses the
most stable branch, masking the hysteresis. It is in this spirit we present
our results in the following.      

Using equation (\ref{eqn5.4}), we have that either 
\begin{equation}
\label{new_eq03}
v_p=v_n(1-S_w)\;, \quad \mbox{for $S_w \to (S_{w,\min})^+$\;,} 
\end{equation}
or 
\begin{equation}
\label{new_eq03-1}
v_p=v_wS_w \quad \mbox{for $S_w \to (S_{w,\max})^-$\;,} 
\end{equation} 
Hence, we have 
\begin{equation}
\label{new_eq03-2}
\frac{dv_p}{dS_w}=-v_n \quad \mbox{for $S_w \to (S_{w,\min})^+$\;,} 
\end{equation}
or 
\begin{equation}
\label{new_eq03-3}
\frac{dv_p}{dS_w}=v_w \quad \mbox{for $S_w \to (S_{w,\max})^-$\;,} 
\end{equation}
Combining these two equations with equation (\ref{eqn19}), we find that
\begin{equation}
\label{eq03-4}
v_m=0  \quad \mbox{for $S_w\to (S_{w,\min})^+$ or $S_w\to (S_{w,\max})^-$\;.}
\end{equation}

We show in figure \ref{fig10}(a) and (b) the wetting and non-wetting seepage velocities as
a function of $S_w$ for pressure gradients $\Delta P/L=0.22$, 0.30,
0.40 and 0.50 MPa/m. The viscosities were $\mu_w=0.03$ Pa s and $\mu_n=0.01$ Pa s.
Both of the seepage velocities $v_w$ and $v_n$ signal a non-zero 
$S_{w,\min}(\Delta P/L)$. However, we find that $S_{w,\max}(\Delta P/L)=1$.  

We denote $v_n=v_n^*$ the non-wetting seepage velocity we find for $S_w < S_{w,\min}(\Delta P/L)$.
It is possible to reach such saturations by initiating the network with a saturation $S_w$ and
a pressure difference $\Delta P_i/L$ making $S_w>S_{w,\min}(\Delta P_i/L)$, and then
reduce the pressure difference to $\Delta P/L$ such that $S_w < S_{w,\min}(\Delta P/L)$.   

We show in figure \ref{fig10}(c) $v_p/v_n$ as a function of $S_w$. The straight line is the 
function $1-S_w$.  By comparing with figure \ref{fig10}(b) that as soon as 
$S_w < S_{w,\min}(\Delta P/L)$, the data for $v_p/v_n$ follows the line $1-S_w$. This is in
accordance with equation (\ref{new_eq03}).    

This teaches us the following: For $v_p=(1-S_w) v_n^*$, i.e., when $S_w < S_{w,\min}(\Delta P/L)$,
we have $dv_p/dS_w=-v_n^*$ and $v_m=0$, see equation (\ref{eq03-4}). If we now compare with figure 
\ref{fig4}, we see that the fits to equation (\ref{eq07}) do not pass through this point,
$(dv_p/dS_w,v_m)=(-v_n^*,0)$.  The difference is too large to be attributed to the uncertainty of
the fits.  We note that we are here dealing with single phase flow. If the constitutive law
for $v_m$ in equation (\ref{eq07}) is the result of correlations appearing in two-phase flow, there
is no reason for the single fluid case to fall on this curve.  

We will in the future present an full analysis of this problem, taking hysteresis fully into account.    

We now turn to the limit where the capillary number is so high that the capillary forces 
are negligible compared to the viscous forces.  We achieve this limit in the dynamic pore
network model by setting the surface tension $\gamma$ in equation (\ref{eqnpc}) to zero.  
If the viscosities of the two fluids are equal, there will be no difference between the 
fluids and $v_w=v_n$. Furthermore, we will have that $v_p$ is independent of the wetting
saturation $S_w$, so that $dv_p/dS_w=0$.  From equation (\ref{eqn19}) we then have that the 
co-moving velocity $v_m=0$.   
  
\begin{figure}[ht]
\centering
\centerline{\includegraphics[width=0.4\textwidth,clip]{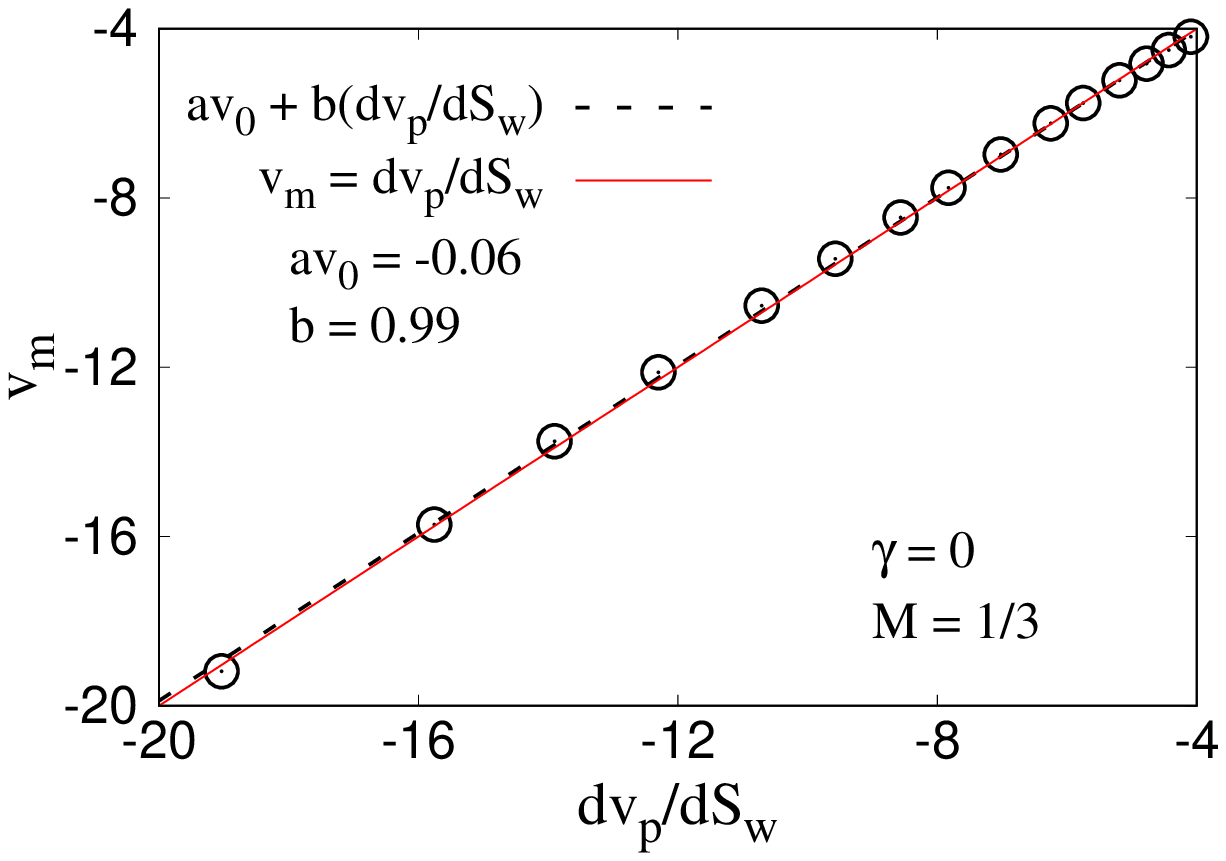}}
\caption{The figure shows how $v_m$ varies with $v_p^{\prime} (=dv_p/dS_w)$ in the limit 
of large capillary numbers when the capillary forces are vanishingly small compared 
to the viscous forces. We have set the surface tension $\gamma=0$ in the dynamic pore 
network model, while keeping $\mu_w=0.03$ Pa s and $\mu_n=0.01$ Pa s.
For reference, the red solid line represents the equation: $v_m=dv_p/dS_w$.}
\label{fig11}
\end{figure}

We show in figure \ref{fig11}, $v_m$ as a function of $dv_p/dS_w$ in the limit of $\gamma=0$
but with the fluid viscosities being $\mu_w=0.03$ Pa s and $\mu_n=0.01$ Pa s respectively. 
We find that $v_m$ follows equation (\ref{eq07}) with $av_0=-0.06$ and $b=0.99$.  From equations
(\ref{eqn5.4}), (\ref{eqn0.3010}) and (\ref{eqn0.3011}), we then have that $v_n=v_w=v_p$.

\subsection{Viscosity ratio $M$}
\label{viscosity}

We will here discuss how the viscosity of the two fluids will affect the relation between $v_p$ and $v_m$. 
We will also discuss how the parameters $av_0$ and $b$ depend on the fluid viscosities $\mu_w$ and $\mu_n$. 

\begin{figure}[ht]
\includegraphics[width=0.4\textwidth,clip]{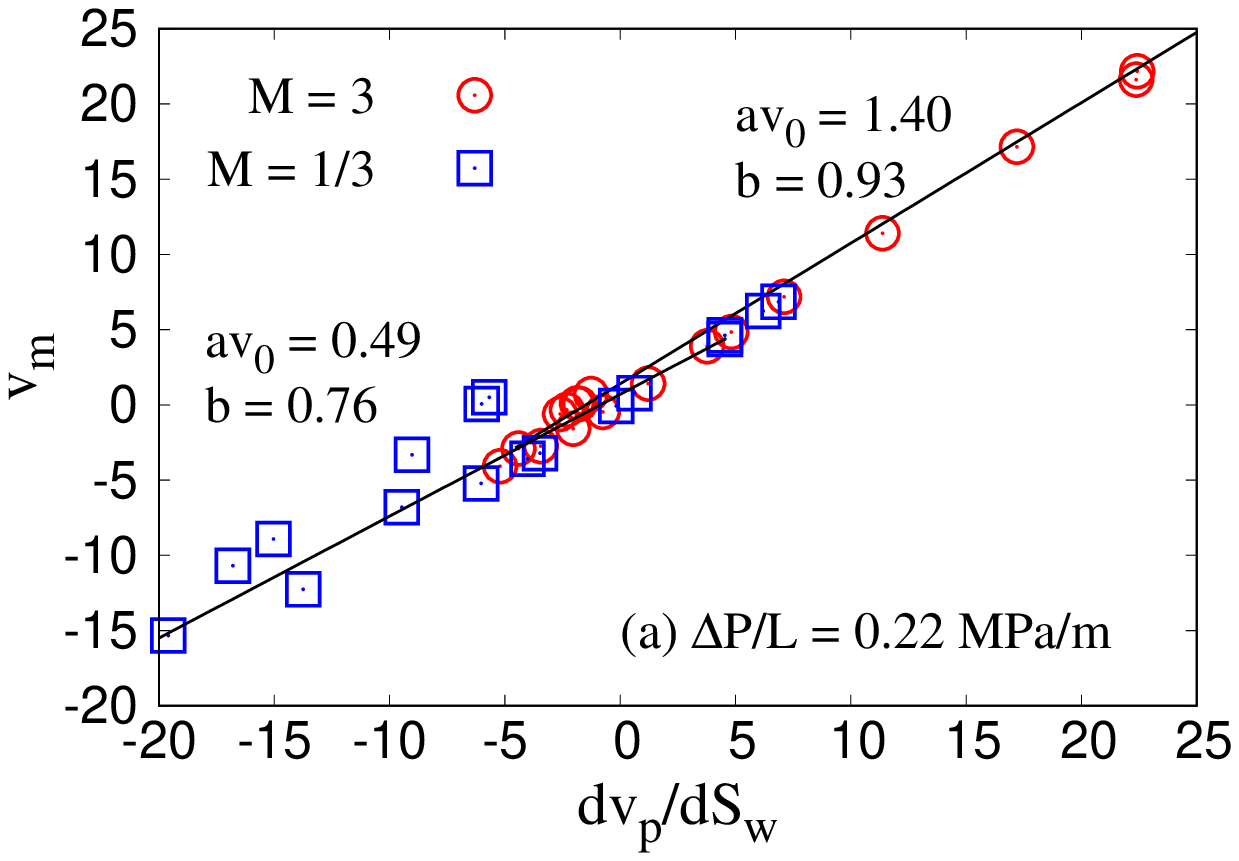} 
\includegraphics[width=0.4\textwidth,clip]{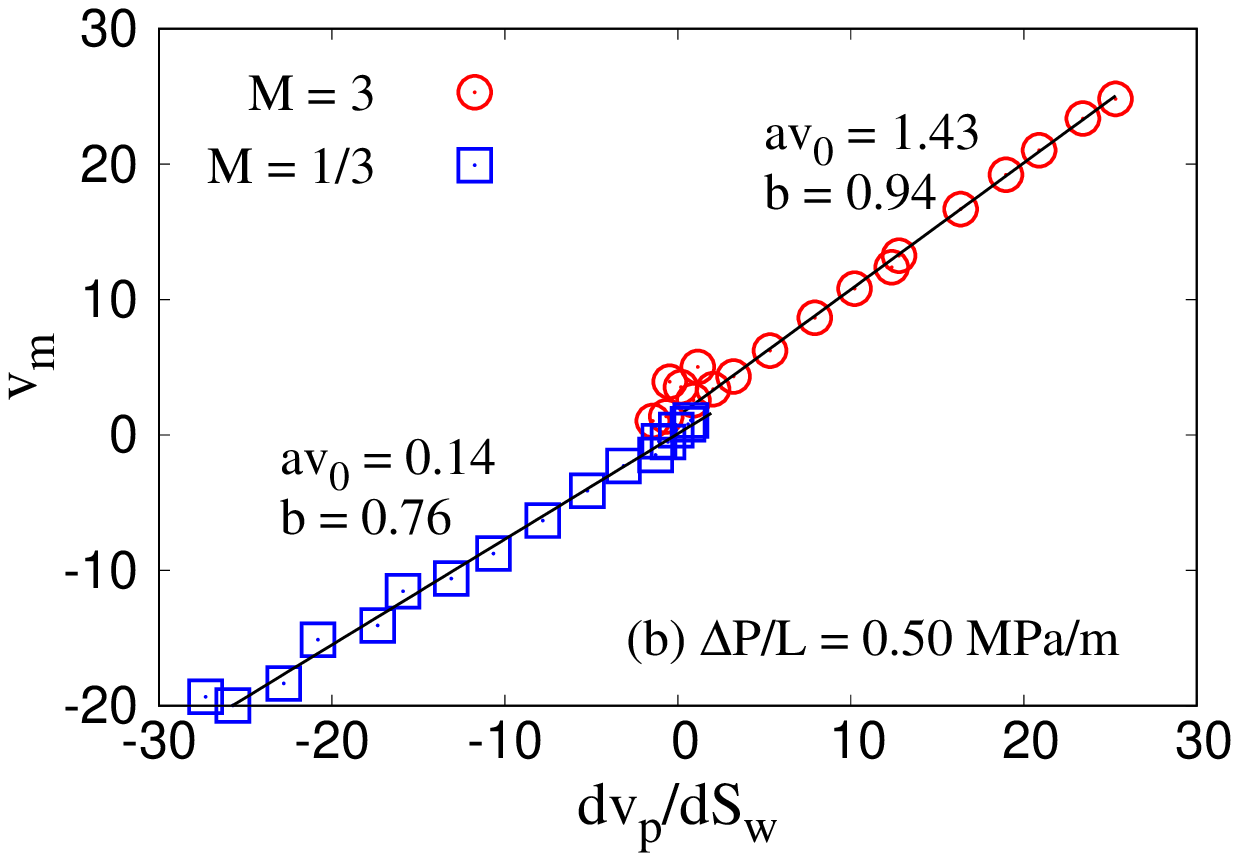} \\ 
\includegraphics[width=0.4\textwidth,clip]{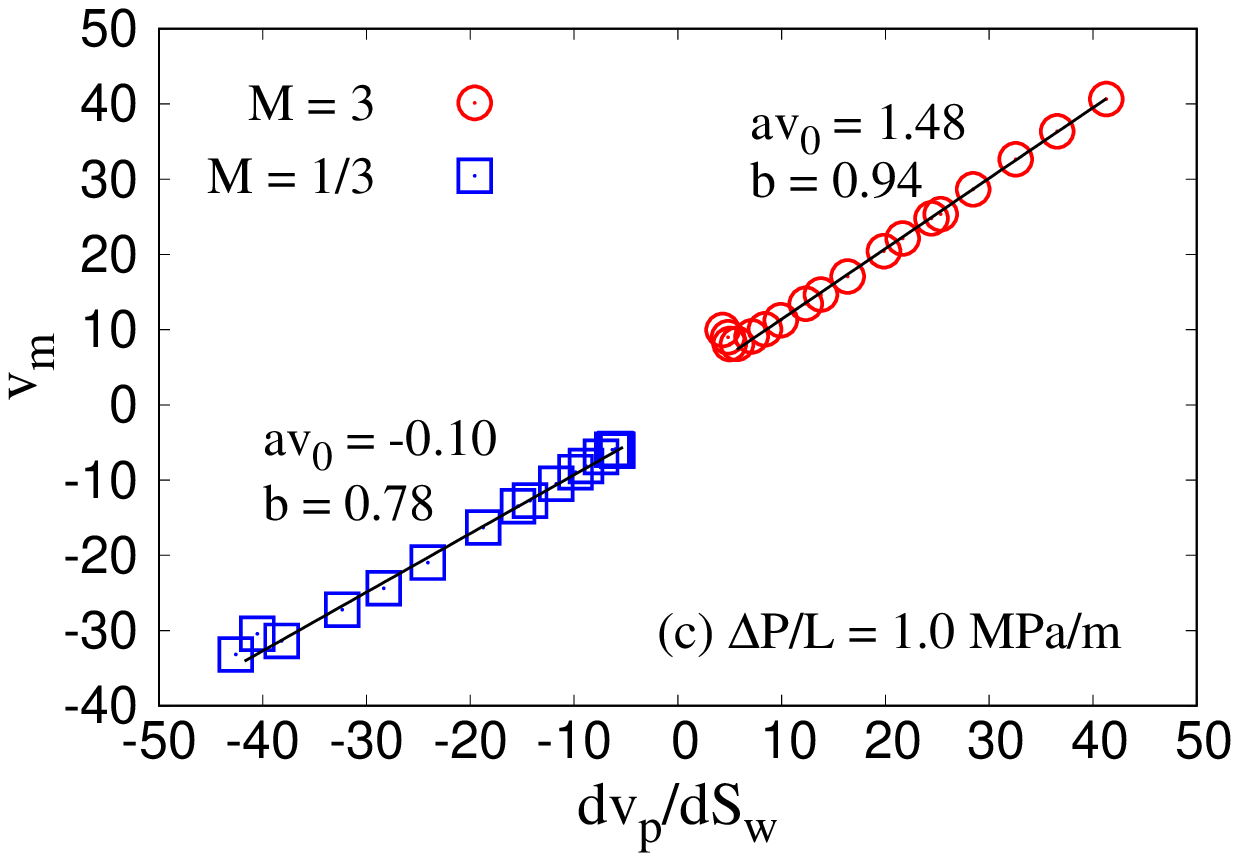} 
\includegraphics[width=0.4\textwidth,clip]{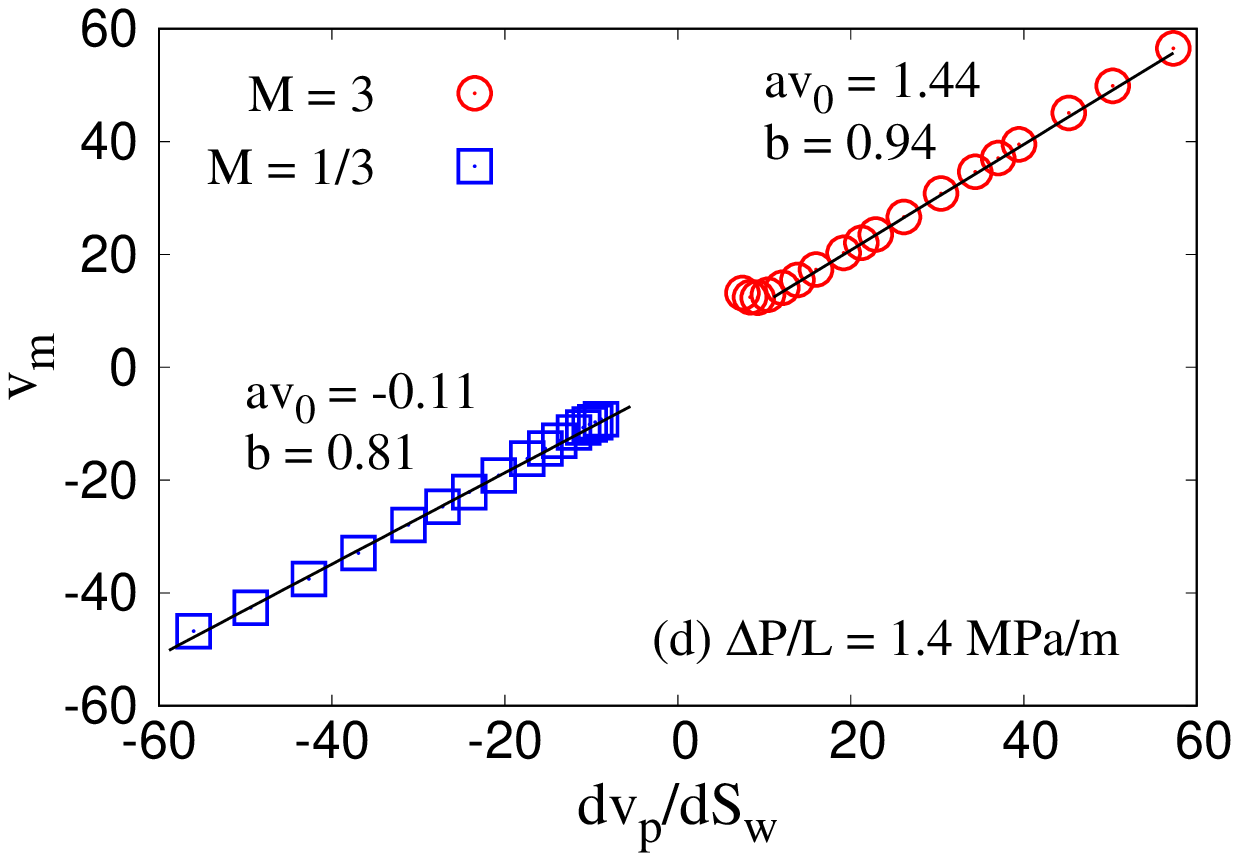}
\caption{Variation of $v_m$ with $dv_p/dS_w$ for four different pressure gradient: 
(a) 0.22, (b) 0.50, (c) 1.0 and (d) 1.4 MPa/m. Results are shown for two different 
viscosity ratios: $M=3$ ($\mu_w=0.03$ Pa s and $\mu_n=0.01$ Pa s) shown as blue squares 
and $M=1/3$ ($\mu_w=0.01$ Pa s and $\mu_n=0.03$ Pa s) as red circles.} 
\label{fig8}
\end{figure}

Figure \ref{fig8} shows how co-moving velocity behaves as a function of $dv_p/dS_w(S_w,\Delta P/L)$ 
for different values of saturation, see equation (\ref{eq07}) when the fluid viscosities are
changed. We compare $v_m$ as a function of $dv_p/dS_w$ for viscosity ratio $M=3$ 
($\mu_w=0.03$ Pa s and $\mu_n=0.01$ Pa s) with viscosity ratio $M=1/3$ ($\mu_w=0.01$ Pa s 
and $\mu_n=0.03$ Pa s).  Figures \ref{fig8}(a), (b), (c) and (d) respectively are based on pressure gradients  
$\Delta P/L$ = 0.22, 0.5, 1.0 and 1.4 MPa/m.  We find that both coefficients $av_0$ and $b$ change
considerably when the viscosity ratio is inverted.  For both $M$ values, 
the co-moving velocity follows equation (\ref{eq07}). For $M= 1/3$, $av_0$ decreases with 
increasing pressure gradient. The coefficient $b$ remains at value around $0.76$ until the 
pressure gradient exceeds a value around $\Delta P/L > 0.5$ MPa/m. On the other hand, for $M=3$, $av_0$ 
and $b$ remain constant around 1.5 and 0.94 respectively irrespective of the pressure gradients we have 
considered. 

\begin{figure}[ht]
\includegraphics[width=0.4\textwidth,clip]{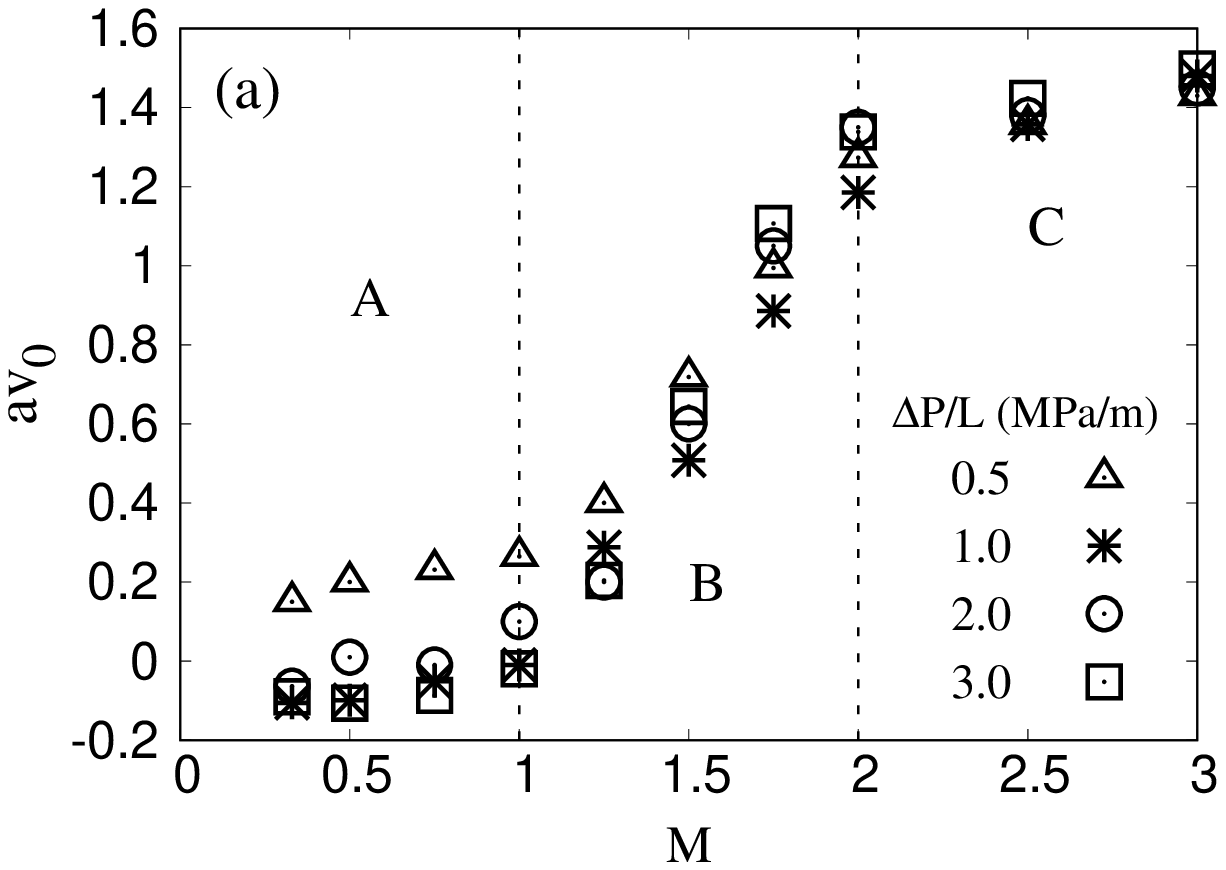} 
\includegraphics[width=0.4\textwidth,clip]{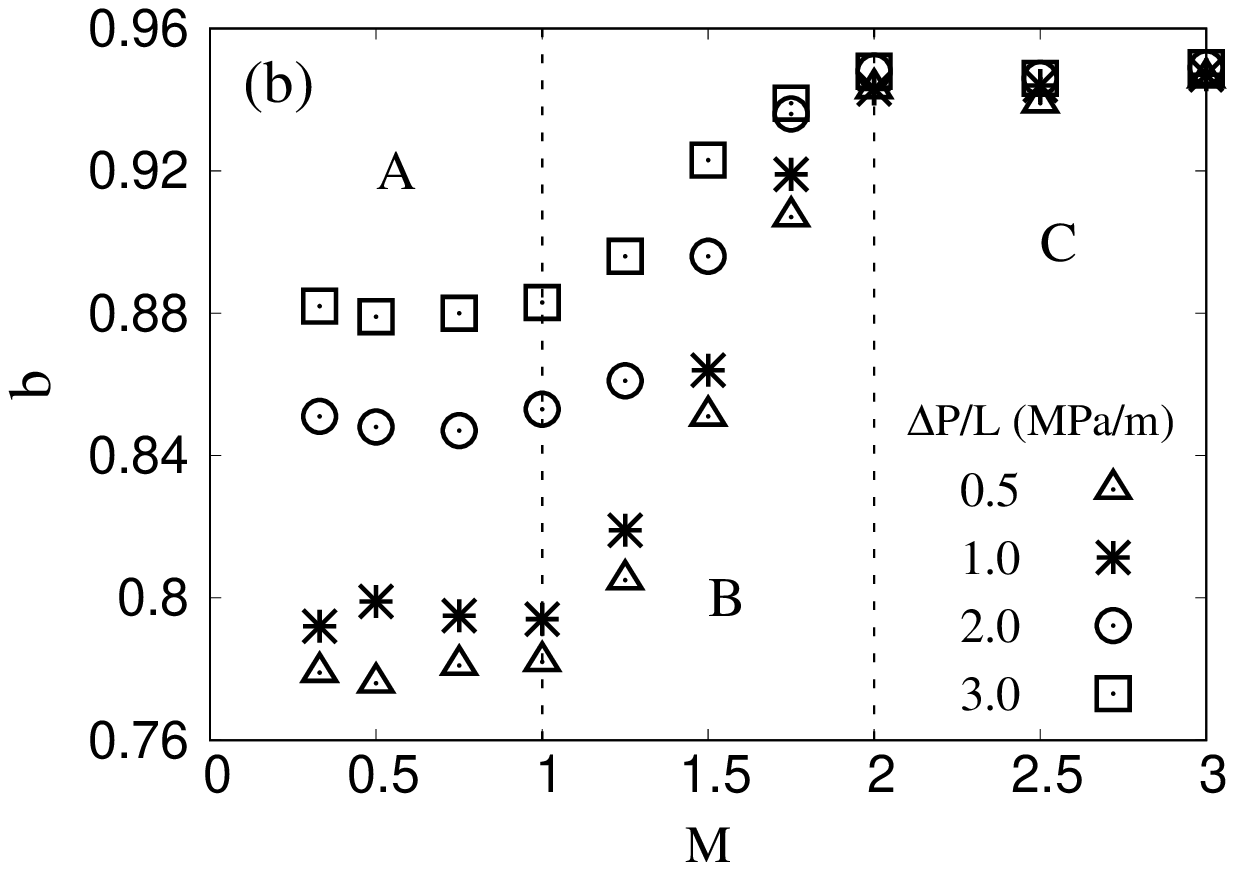} 
\caption{Variation of $av_0$ and $b$ with viscosity ratio $M$ for a constant pressure gradient 
$\Delta P/L$. The numerical results are repeated for four different pressure gradients, 
$\Delta P/L= 0.5$, 1.0, 2.0 and 3.0 MPa/m. For $M=1.0$, we set $\mu_n=0.01$ Pa s and $\mu_w=0.01$ Pa s. 
For $M>1$, we keep $\mu_w=0.01$ Pa s while $\mu_n$ has a value $M\mu_w$ Pa s. On the other hand, for $M<1$, 
we keep $\mu_n=0.01$ Pa s while $\mu_w$ has a value $\mu_n/M$ Pa s.}
\label{fig9}
\end{figure}

We plot in figure \ref{fig9} the coefficients $av_0$ and $b$ as functions of the viscosity
ratio $M$ for different values of the pressure gradient $\Delta P/L$. For the chosen span of
parameters, we observe three distinct regions:

\noindent
{\sl Region A ($M \le 1$)} - In this region, both $av_0$ and $b$ seem independent of $M$. Moreover, $b$ 
remains constant around 0.77 for $\Delta P/L < 0.5$ MPa/m and increases for larger values of the 
pressure gradient. $av_0$ decreases with increasing pressure gradient as long as $\Delta P/L < 0.5$ MPa/m. 
Beyond this limit, $av_0$ saturates at a value close to zero.    

\noindent
{\sl Region B ($1 \le M \le 2$)} - In this region $av_0$ and $b$ are both increasing functions of $M$. 

\noindent
{\sl Region C ($M \ge 2$)} - In this region, $av_0$ and $b$ neither changes with viscosity ratio $M$ 
nor with the pressure gradient $\Delta P/L$. 

\section{Discussion}
\label{discussion}

The aim of this paper has been to expand on the theory based on Euler homogeneity 
that was first presented in \cite{hsbkgv18}.  It provides a number of relations 
between the seepage velocities of each fluid involved which together with 
constitutive equations for the average fluid velocity and the co-moving 
velocity form a closed set of equations. 

It has recently been discovered that the 
constitutive equation for the average seepage velocity of the fluids follows a power law 
in the pressure gradient for a range of parameter values 
\cite{tkrlmtf09,tlkrfm09,aetfhm14,sh17,glbb20,zbglb21}.
Relative permeability theory offers the mapping  $(v_w,v_n)\to v_p$.  However, the
non-linear constitutive equation for $v_p$ requires the opposite mapping 
$v_p\to(v_w,v_n)$, which is indeterminate within relative permeability theory. Euler
homogeneity theory, on the other hand, offers the two-way mapping 
$(v_w,v_n)\leftrightarrows (v_p,v_m)$,
which is readily combined with the non-linear constitutive equation for $v_p$.
It is an additional bonus that the constitutive equation for $v_m$, 
equation (\ref{relperm-200}), is as simple as it is. 

The co-moving velocity which together with the average seepage velocity of the fluids closes the
equation set as described in Section \ref{closed}, is related to the seepage velocity
difference $v_n-v_w$, but it is not the same, see equations (\ref{eqn19}) and (\ref{eqn20}).
We discuss in Section \ref{interpreting} the interpretation of $v_m$.  It should be 
noted that the co-moving velocity is not associated volume transport, see equation (\ref{euler-22}).

We determine the constitutive equation for the co-moving velocity from relative permeability
data found in the literature in Section \ref{expResults}. We do this by {\it reverse engineer\/}
the data which have been cast in the form of relative permeability curves.      

Our main result is equation 
(\ref{relperm-200}), which shows that the co-moving velocity is linear in the derivative
of the average seepage velocity with respect to the saturation when the pressure gradient is 
kept fixed, see figures \ref{fig:Bennion_experimental} to \ref{fig:Leverett_experimental}.  
It is an open question as to why this is so. 

Since we do not have  theory as to why the co-moving velocity takes the simple form it
does, we have not applied any particular criterion for which data sets to investigate. 
Any attempt at this would taint the results by our preconception on what causes 
the functional form (\ref{relperm-200}).  In particular, we have not taken the
possibility for hysteresis into account.  We discuss why it is still permissible to
treat hysteretic data as representatives of analytic functions in Subsection \ref{hysteresis}. 

We continue in Section \ref{poreNetwork} to consider the constitutive equation for the 
co-moving velocity.  We find the same constitutive equation as in equation 
(\ref{relperm-200}), see figure \ref{fig4}. It is remarkable that this remains true also 
when the constitutive equation for the average seepage velocity moves into the power-law region,
see Section \ref{vp_nonlinear}.

We have in this paper only considered systems without a saturation gradient.  This has
allowed us to ignore capillary pressure effects.  A next step is to incorporate such a 
saturation gradient into the system to observe how the constitutive equation (\ref{relperm-200})
for the co-moving velocity changes.


\section{Acknowledgment}
This work was partly supported by the Research Council of Norway through its Centres
of Excellence funding scheme, project number 262644. SS was supported
by the National Natural Science Foundation of China under grant number
11750110430. We thank D.\ Bedeaux, C.\ F.\ Berg, H.\ Cheon, H.\ Fyhn, M.\ Aa.\ Gjennestad, 
S.\ Kjelstrup and P.\ A.\ Slotte for discussions.

\bigskip


\end{document}